\documentclass[aps,prd,showpacs,twocolumn,floatfix,nofootinbib,superscriptaddress,
amsmath,amssymb,amsfonts,]{revtex4-1} 

\usepackage{bm}
\usepackage{slashed}
\usepackage{graphicx}

\usepackage{amsmath,amssymb,amsfonts,graphicx}
\usepackage{bm}
\usepackage{slashed}
\usepackage{subfigure}
\usepackage[svgnames]{xcolor}
\usepackage{placeins}
\usepackage{array}
\usepackage{natbib,hyperref}  
\usepackage{url}
\usepackage[utf8]{inputenc}

\makeatletter
\newcommand{\thickhline}{%
    \noalign {\ifnum 0=`}\fi \hrule height 1pt
    \futurelet \reserved@a \@xhline
}
\newcolumntype{"}{@{\hskip\tabcolsep\vrule width 1pt\hskip\tabcolsep}}
\makeatother

\usepackage{placeins}

\usepackage{xspace}

\newcommand{\Eq}[1]{Eq.~(\ref{#1})}
\newcommand{\al}[1]{\begin{align} #1 \end{align}}
\newcommand{\non}{\nonumber}
\newcommand{\vect}[1]{\boldsymbol{#1}}

\def\vf{\varphi}

\def\pp#1{\vect{p}_{#1\perp}}

\newcommand{\psq}[1]{p_{#1\perp}^2}

\newcommand{\pe}[1]{\vect{p}_{{#1}\perp}}

\def\sT{\vect{s}_T}

\newcommand{\sq}[1]{\vect{s}_{#1}}

\newcommand{\sqt}[1]{\vect{s}_{{#1}T}}

\newcommand{\Gs}{\mathrm{GeV}^2}
\newcommand{\ImM}{0.85\columnwidth}
\newcommand{\ImL}{0.85\columnwidth}
\newcommand{\GapCapt}{\vspace{-0pt}}
\newcommand{\GapSubf}{\vspace{-8pt}}
\begin{document}

\title{Monte Carlo Implementation of Polarized Hadronization}

\preprint{ADP-16-36/T992}

\author{Hrayr~H.~Matevosyan}
\thanks{ORCID: http://orcid.org/0000-0002-4074-7411}
\affiliation{ARC Centre of Excellence for Particle Physics at the Tera-scale,\\ 
and CSSM, School of Chemistry and Physics, \\
The University of Adelaide, Adelaide SA 5005, Australia
\\ http://www.physics.adelaide.edu.au/cssm
}

\author{Aram~Kotzinian}
\thanks{ORCID: http://orcid.org/0000-0001-8326-3284}
\affiliation{Yerevan Physics Institute,
2 Alikhanyan Brothers St.,
375036 Yerevan, Armenia
}
\affiliation{INFN, Sezione di Torino, 10125 Torino, Italy
}

\author{Anthony~W.~Thomas}
\thanks{ORCID: http://orcid.org/0000-0003-0026-499X}
\affiliation{ARC Centre of Excellence for Particle Physics at the Tera-scale,\\     
and CSSM, School of Chemistry and Physics, \\
The University of Adelaide, Adelaide SA 5005, Australia
\\ http://www.physics.adelaide.edu.au/cssm
}

\begin{abstract}
 We study the polarized quark hadronization in a Monte Carlo (MC) framework based on the recent extension of the quark-jet framework, where a self-consistent treatment of the quark polarization transfer in a sequential hadronization picture has been presented. Here, we first adopt this approach for MC simulations of the hadronization process with a finite number of produced hadrons, expressing the relevant probabilities in terms of the eight leading twist quark-to-quark transverse-momentum-dependent (TMD) splitting functions (SFs) for elementary $q \to q'+h$ transition. We present explicit expressions for the unpolarized and Collins fragmentation functions (FFs) of unpolarized hadrons emitted at rank 2. Further, we demonstrate that all the current spectator-type model calculations of the leading twist quark-to-quark TMD SFs violate the positivity constraints, and we propose a quark model based ansatz for these input functions that circumvents the problem.  We validate our MC framework by explicitly proving the absence of unphysical azimuthal modulations of the computed polarized FFs, and by precisely reproducing the earlier derived explicit results for rank-2 pions. Finally, we present the full results for pion unpolarized and Collins FFs, as well as the corresponding analyzing powers from high statistics MC simulations with a large number of produced hadrons for two different model input elementary SFs. The results for both sets of input functions exhibit the same general features of an opposite signed Collins function for favored and unfavored channels at large $z$ and, at the same time, demonstrate the flexibility of the quark-jet framework by producing significantly different dependences of the results at mid to low $z$ for the two model inputs.
\end{abstract}



\date{\today}                                           

\maketitle

\section{Introduction}
\label{SEC_INTRO}
 
 One of the more fascinating topics in modern high energy physics is the description of hadronization of partons after hard scattering. Within a QCD factorized approach, this part of the inclusive cross section is described by nonperturbative parton fragmentation functions (FFs)~\cite{Metz:2016swz}. These were introduced 40 years ago by Field and Feynman~\cite{Field:1976ve, Field:1977fa}. For each parton flavor $q$ and hadron type $h$, they are a function of the ratio of the hadron light-cone momentum to the fragmenting quark light-cone momentum $z$, $D_q^h(z)$. In this simplest form, the polarization of partons and the produced hadron transverse momentum (with respect to the parton's flight direction)  were not considered. These FFs, often called collinear,  have proven to be a powerful tool for studying the nucleon structure. For example, using the data from deep inelastic semi-inclusive hadron production (SIDIS), these one-dimensional FFs allow one to disentangle the quark flavor distributions in an unpolarized or a longitudinally polarized nucleon. 

The complete description of partonic constituents for a fast moving nucleon is encoded in the spin and parton transverse-momentum-dependent (TMD)  parton distribution functions (PDFs). One can access these PDFs using, for example, SIDIS from a polarized nucleon. In particular, the quark transversity PDF can be accessed if the TMD FF of a transversely polarized quark into an unpolarized hadron, the so-called Collins FF, does not vanish~\cite{Collins:1992kk}. At present, both the collinear and TMD FFs are parametrized by fitting high energy inclusive data; recent phenomenological fits of the pion Collins function can be found in Refs.~\cite{Kang:2015msa,Anselmino:2015sxa}. To better understand the physics of hadronization, several dynamic models, such as the quark-diquark spectator model~\cite{Bacchetta:2007wc} and the quark-jet model~\cite{Ito:2009zc} based on Nambu--Jona-Lasinio (NJL) effective quark theory~\cite{Nambu:1961tp,Nambu:1961fr}, have been used. Recently, important progress has been achieved in modeling polarized hadronization in the quark-jet approach~\cite{Bentz:2016rav}. In this work all eight elementary polarized TMD $q \to q'$ splitting functions, which can be calculated by using effective quark theories, were taken into account in the resulting integral equations for polarized TMD FF.

Another widely used approach describing hadronization is based on the Lund string model~\cite{Andersson:1983ia} and implemented in the Monte Carlo event generators PYTHIA~\cite{Sjostrand:2007gs} and LEPTO~\cite{Ingelman:1996mq}. In these event generators it is rather straightforward to include the polarization of the initial nucleon and the active parton. Then it becomes possible to include, in these MC programs, some of the spin-dependent effects, like the Sivers effect\cite{Sivers:1989cc} in SIDIS, and describe the existing data and make predictions for forthcoming experiments, as was done in Refs.~\cite{Kotzinian:2014lsa,Kotzinian:2014gza,Matevosyan:2015gwa}. On the other hand, because at present there is no MC framework for polarized parton hadronization, it is not possible to simulate the Collins effect~\cite{Collins:1992kk}, which appears to be significant in SIDIS and electron-positron annihilation measurements. A proposal on how to include quark transverse polarization effects in hadronization was presented in~\cite{Artru:2014abc}, but still no MC realization of this framework exists. Finally, it is important to note that there is no clear way to extract the independent quark fragmentation functions from the Lund string model, where the hadronization of the color flux string depends on both the type of the initial quark of interest at one end of the string and the colored remnant (antiquark in $e^+ e^-$, diquark in SIDIS, etc) at the other end~\cite{Kotzinian:2004xq}.

\begin{figure}[b]
\centering 
\includegraphics[width=\ImM]{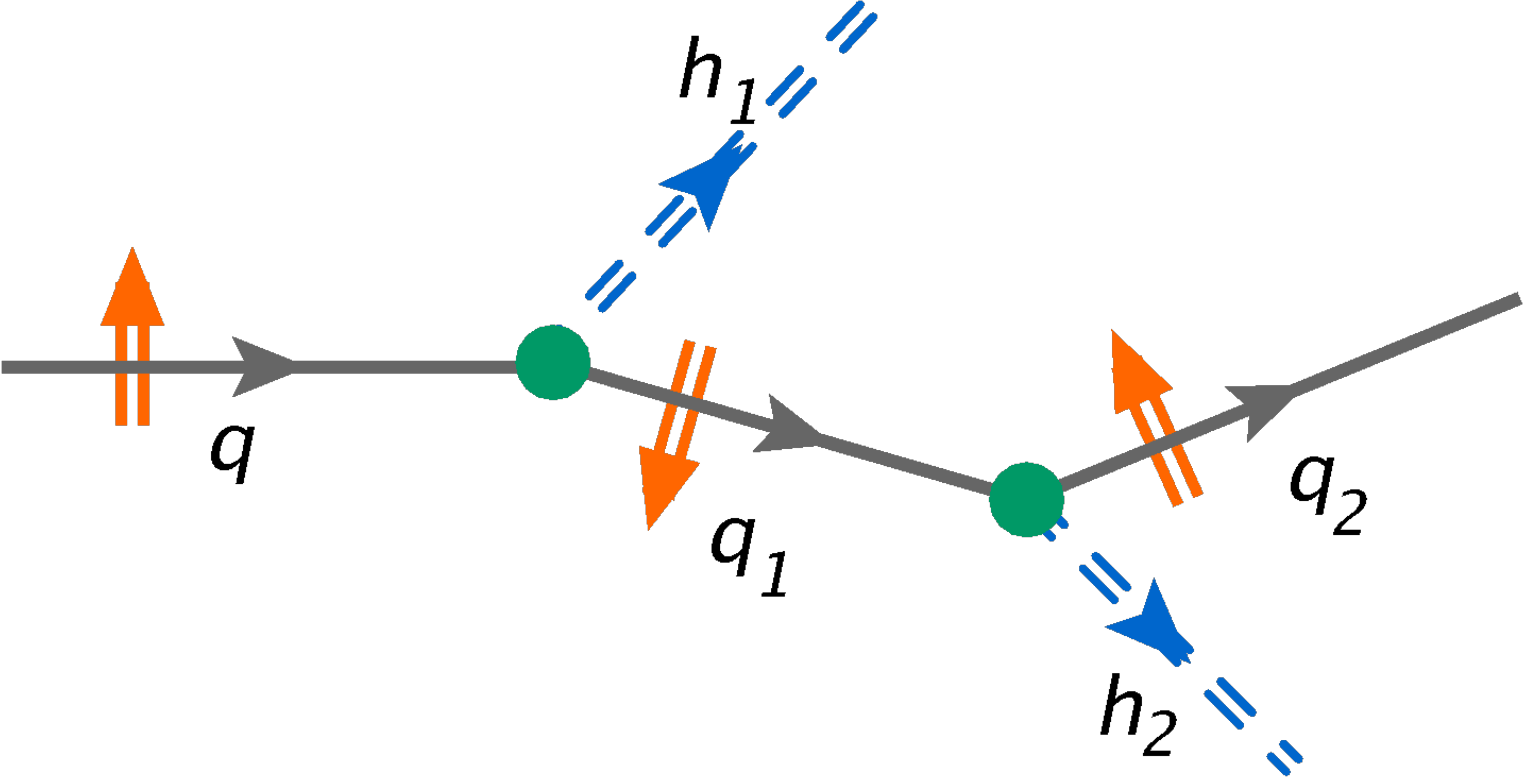}
\GapCapt
\caption{Schematic depiction of the extended quark-jet framework.}
\label{PLOT_QUARK-JET}
\end{figure}

In this paper we describe the MC framework for transversely polarized quark to pion FFs based on the extended quark-jet framework of Bentz {\it et al.}~\cite{Bentz:2016rav}. The quark-jet framework describes the hadronization of a quark as a sequential emission of hadrons that do not interact with each other or re-interact with the remnant, as schematically depicted in Fig.~\ref{PLOT_QUARK-JET}. The quark to hadron fragmentation functions are then calculated as the corresponding number densities, either using integral equations or Monte Carlo techniques. The original model of Field and Feynman~\cite{Field:1976ve,Field:1977fa} has been significantly extended in recent years to describe various phenomena in hadronization in the so-called NJL-jet model, which  uses the NJL effective quark model~\cite{Nambu:1961tp,Nambu:1961fr} to calculate the input elementary hadron emission probabilities. The extensions include the calculations of the collinear FFs for various hadrons~\cite{Ito:2009zc, Matevosyan:2010hh,Matevosyan:2011ey,PhysRevD.86.059904}, transverse-momentum-dependent FFs \cite{Matevosyan:2011vj}, dihadron FFs~\cite{Matevosyan:2013nla, Matevosyan:2013aka,Casey:2012ux,Casey:2012hg} and spin-dependent effects~\cite{Matevosyan:2012ga,Matevosyan:2012ms,Matevosyan:2013eia}. The latter have proven especially challenging, as the naive interpretations of the polarization transfer dynamics lead to higher-order Collins modulations~\cite{Matevosyan:2012ed} that are nonphysical, while the probabilities of hadron emission should only depend linearly on the polarization of the initial quark. This problem  was circumvented in Ref.~\cite{Matevosyan:2013eia} by including only a single emission step with Collins modulation, which allowed one to study the connection between polarization induced azimuthal modulations in one- and two-hadron FFs recently observed in the COMPASS experiment~\cite{Adolph:2014fjw}. 
 
 This paper is organized in the following way.  In the next section we present the theoretical framework behind the MC generator. In Sec.~\ref{SEC_SPLITTINGS} we briefly describe the details of the model used to extract the polarized FFs. In Sec.~\ref{SEC_RESULTS}  we present the MC computations for the unpolarized and Collins FFs of pions produced by an up quark, and we finish with the conclusions in Sec.~\ref{SEC_CONCLUSIONS}.

\section{Quark-jet framework and the polarization transfer}
\label{SEC_THEORY}

 Recently, we have derived a general, self-consistent formalism within the quark-jet framework~\cite{Bentz:2016rav} to describe the hadronization of a polarized quark that is independent of the details of the model input splitting functions. Here we reiterate some of the key points in this derivation and adapt it for MC simulations.

\subsection{Intermediate quark polarization}
\label{SUBSEC_INT_QUARK_SPIN}

  The key component in building the extended quark-jet model is the description of the polarization of the remnant quark in the jet after each  hadron emission. Here, to describe  this process, we use the spin density matrix formalism of Ref.~\cite{Berestetsky:1982aq}, which has been successfully used in the past for describing the polarized SIDIS cross section, e.g.~\cite{Kotzinian:1994dv}.  
    
  In general, the polarization of a spin $1/2$ particle $q$ is describe by the spin density matrix $\rho$, which can be expressed in terms of the Pauli-Lubanski 4-vector $a$,
\al
{
   \rho_q = \frac{1}{2}(\slashed{p} +m ) [1 - \gamma^5 \slashed{a}],
}
where $p$ and $m$ are the 4-momentum and the mass of $q$. The 4-vector $a$ is defined in the particle's rest frame as
\al
{
 a = (0, \sq{q}),
}
where the polarization vector $\sq{q}$ itself equals twice the expectation value of the spin of the particle at rest.

 Let us consider the elementary FF in the first fragmentation step $q\to q_1$. The probability density for this transition can be expressed in terms of the respective density matrices $\rho_{q}$ and $\rho_{q_1}$,
 \al
{
  f^{q\to q_1} = Tr[\rho_{q_1} A \rho_q \bar{A}],
}
where $A$ is some matrix describing the interaction with the other particles in this process. It is more convenient to work with the corresponding polarization vectors $\sq{q}$ and $\sq{q_1}$. Then the probability density $f^{q\to q_1}$ should be a linear function in both $\sq{q}$ and $\sq{q_1}$, 
\al
{
\label{EQ_EL_FF_SPIN}
 f^{q\to q_1}( \vect{s}_q, \sq{q_1}) = \alpha_q + \vect{\beta}_q \cdot \sq{q_1},
}
where  both $\alpha_q$ and $\vect{\beta}_q$ are linear functions of $\vect{s}_q$ that also depend on the momenta of the quarks. We can express these coefficients in terms of the $8$ leading-twist  quark-to-quark  TMD SFs [see Eq.~(2.19) in Ref.~\cite{Bentz:2016rav}] 
 \al
{
\label{EQ_EL_ALPH_BETA}
	\alpha_q \equiv &   \hat{D}(z_1, \psq{1}) +  \frac{ (\pp{1} \times \sT)\cdot {\vect{\hat{z}}}}{z_1\mathcal{M}} \ \hat{H}^\perp(z_1,\psq{1}),
\\
	{\beta}_{q \parallel}  \equiv  &  s_L \ \hat{G}_L(z_1,\psq{1})  -  \frac{ (\pp{1} \cdot \sT)}{z_1\mathcal{M}} \hat{H}_L^\perp(z_1,\psq{1}),
\\
	\vect{\beta}_{q \perp} \equiv  & \frac{ \pp{1}' }{z_1 \mathcal{M}}  \hat{D}_T^\perp(z_1,\psq{1})  - \frac{\pp{1} }{z_1 \mathcal{M}} s_L  \hat{G}_T(z_1,\psq{1})
\\\non
&+\sT \ \hat{H}_T(z_1,\psq{1})  + \frac{\pp{1} (\pp{1} \cdot \sT)}{z_1^2 \mathcal{M}^2} \ \hat{H}_T^\perp(z_1,\psq{1}), 
}
where $z_1$ and $\pp{1}$  are the light-cone momentum fraction and the transverse momentum of $q_1$ with respect to $q$, while $\mathcal{M}$ is the mass of $q_1$. The momentum  vector $\pp{1}' \equiv (-p_{1,y}, p_{1,x})$. The unit vector $\hat{\vect{z}}$ denotes the direction of the 3-momentum of $q$, which also helps to define $\sT$ and $s_L $  as the transverse and longitudinal components of $\sq{q} = ( \sT,s_L)$. In this work we use hats on TMD SFs to distinguish them from the analogous TMD FFs . 

In the extended quark-jet model this quark $q_1$ itself fragments into a hadron and a remnant quark $q_2$. Indeed, the quark $q_1$ is unobserved, and its polarization is completely determined by $\sq{q}$, $z_1$ and $\pp{1}$. According to Ref.~\cite{Berestetsky:1982aq}, it can be expressed as
\al
{
\label{EQ_REMNAT_SPIN}
  \sq{q_1} = \frac{\vect{\beta}_q}{\alpha_q}.
}

The probability to produce quark $q_1$ with light-cone momentum fraction $z_1$ and transverse momentum $\pp{1}$ is determined from \Eq{EQ_EL_FF_SPIN},
\al
{\label{EQ_PROB_UNPOL}
  \hat{f}^{q\to q_1}(z_1, \pp{1}; \sq{q}) = \alpha_q.
}

 Then for the next fragmentation step $q_1 \to q_2$, we have a completely analogous situation, where 
\al
{
  f^{q_1 \to q_2}( \sq{q_1}, \sq{q_2}) = \alpha_{q_1} + \vect{\beta}_{q_1} \cdot \sq{q_2},  
}
and
\al
{
\sq{q_2} = \frac{\vect{\beta}_{q_1}}{\alpha_{q_1}}.
}
Here again, $\alpha_{q_1}$ and $\vect{\beta}_{q_1}$ are both linear functions of $\sq{q_1}$ that also depend on the momentum of $q_2$ with respect to $q_1$. We can write for them analogous relations to those in \Eq{EQ_EL_ALPH_BETA}, involving the light-cone momentum fraction $\eta_2$ and transverse momentum $\vect{p}_{2 \perp}$  of quark $q_2$ relative to $q_1$. Nevertheless, since $\sq{q_1}$ itself is determined by $\sq{q}$, we can infer that $\sq{q_2}$ should also be completely determined by $\sq{q}$, as well as the light-cone momentum fraction $z_2 $ and transverse momentum $\vect{P}_{2\perp }$ of quark $q_2$ with respect to $q$. Then, in the quark-jet framework, the probability of the $q\to q_2$ transition is given by
\al
{
  \hat{f}^{(2) }_{q\to q_2}(z_2, \vect{P}_{2 \perp}; \sq{q})   &
  \\\non 
  = \hat{f}^{q\to q_1}&(z_1, \pp{1}; \sq{q}) \otimes \hat{f}^{q_1 \to q_2}(\eta_2, \pp{2}; \sq{q_1}),
}
where 
\al
{
   \hat{f}^{q_1 \to q_2}(\eta_2, \vect{p}_{2\perp }; \sq{q_1})  = \alpha_{q_1},	
}
and the convolution $\otimes$ relates the corresponding relative momenta (the detailed relations will be discussed in the next section).
 
 We can conclude that for the remnant quark $q_N$ after $N$ emissions, the polarization $\sq{q_N}$ is completely determined by the momenta of the quarks in the chain and the polarization $\sq{q}$ of the initial fragmenting quark $q$. Within the quark-jet framework, the probability for this quark to have certain momentum with respect to the initial quark is a convolution of elementary probabilities that themselves are determined by the polarization of the initial fragmenting quark and the momenta of all the quarks in the jet up to the one under consideration. These elementary probabilities are those for polarized quark splitting into an unpolarized quark [see \Eq{EQ_PROB_UNPOL}].

\subsection{Monte Carlo approach}
\label{SUBSEC_MC_APPROACH}

The application of the quark polarization propagation mechanism in the quark-jet hadronization chain with an infinite number of produced hadrons results in a set of coupled  integral equations for the unpolarized and Collins FFs, as detailed in Ref.~\cite{Bentz:2016rav}. Also, this iterative picture allows us to readily adapt the extended quark-jet framework for MC simulations with a finite number of produced hadrons, similar to our previous work in Refs.~\cite{Matevosyan:2011vj,Matevosyan:2012ga, Matevosyan:2012ms}. The basic concept  is to adapt the number density implementation of the FFs, which then can be calculated using Monte Carlo techniques as averages of these densities taken over a large number of quark hadronization event simulations. In the instance of polarized quark fragmentation into unpolarized hadrons, the corresponding number density is the following polarized fragmentation function:
\begin{align}
\label{EQ_Dqh_SIN}
D_{h/q^{\uparrow}} (z,\psq{},\varphi) &= D^{h/q}(z,\psq{})
\\ \nonumber
 &- H^{\perp h/q}(z, \psq{}) \frac{ p_\perp s_{T}}{z m_h} \sin(\vf_C),
\end{align}
where $D^{h/q}(z,\psq{})$ and $H^{\perp h/q}(z, \psq{})$ denote the unpolarized  and Collins fragmentation function, respectively. The variables $z$ and $\psq{}$ are the light-cone momentum fraction and the transverse momentum squared of the produced hadron with respect to the momentum of the initial fragmenting quark, and $m_h$ denotes its mass. Here, $s_T$ is the modulus of the transverse component of the quark's polarization. The Collins angle for the hadron $\vf_C \equiv \vf - \vf_s$ is defined as the difference of the azimuthal angles of the produced hadron's transverse momentum $\vf$ and the transverse polarization of the initial quark $\vf_s$. We calculate $D_{h/q^{\uparrow}} (z,\psq{},\varphi) $ by computing the average number of  hadrons $h$ with given momenta produced in the hadronization chain of $q$. This can be accomplished by sampling the remnant quark's momentum according to the elementary quark-to-quark splitting functions,~\Eq{EQ_PROB_UNPOL}, and calculating the type and the momentum of the produced hadron using  flavor and momentum conservation. Equivalently, we can sample the quark-to-hadron splitting functions and reconstruct the remnant quark's type and momentum. The remnant quark's polarization is then determined according to~\Eq{EQ_REMNAT_SPIN}. We can continue the hadronization chain until we reach some predetermined termination condition, which we choose as a given number of produced hadrons $N_L$. In the text below we denote the hadrons produced at the $n$th step in the hadronization chain as rank-$n$ hadrons.
  
\subsection{Two-step process}
\label{SUBSEC_TWO_STEP}

\begin{figure}[htb]
\centering 
\includegraphics[width=\ImL]{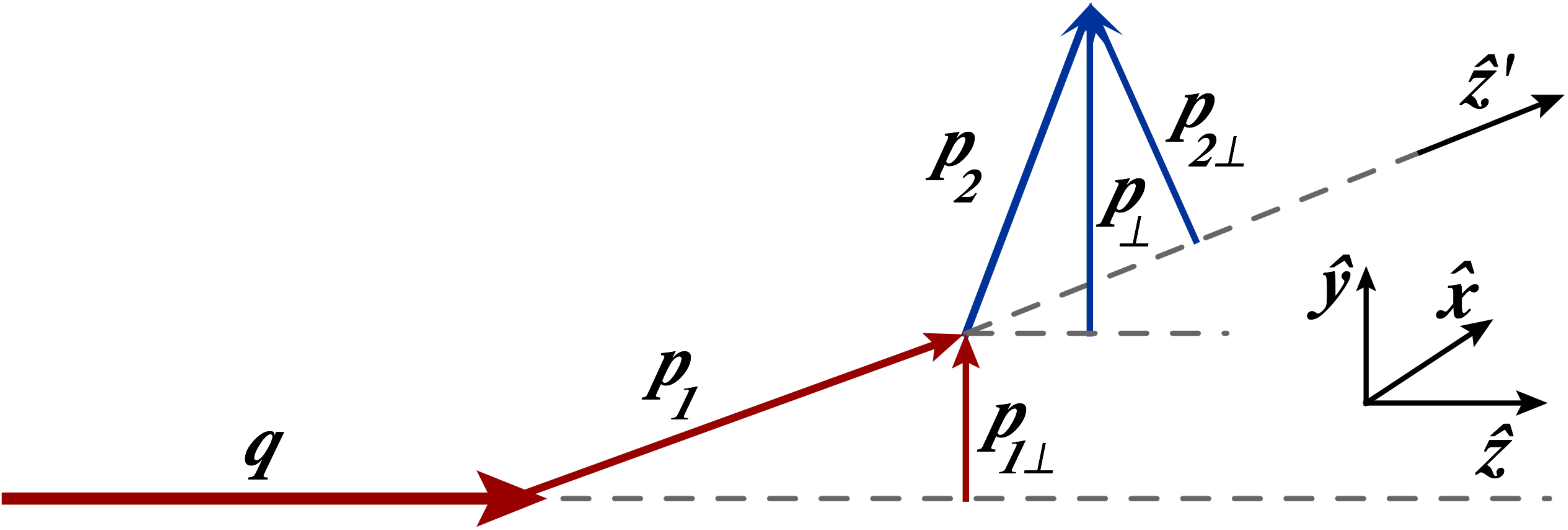}
\GapCapt
\caption{Depiction of the two-step process kinematics in the quark-jet picture. The axis $\hat{\vect{z}}$ is defined by the 3-momentum $\vect{q}$ of the initial quark $q$, while the axis $\hat{\vect{z}}'$ is defined by the momentum vector $\vect{p}_1$  of the first remnant  quark ${q}_1$. The vectors $\vect{p}_2$ and $\vect{p}$ are the momenta of the second produced hadron $h$ in the two different coordinate systems.}
\label{PLOT_Q_to_Q_H}
\end{figure}

 Here we discuss two-hadron production in a quark-jet picture, where the initial quark $q$ emits rank-one hadron $h_1$ with remnant quark $q_1$, which in turn emits the rank-2 hadron $h$ and leaves a remnant quark $q_2$. We are interested in the azimuthal modulations of the transverse momentum of $h$ when the initial quark $q$ has nonzero transverse polarization, denoted as $\sT$. Hence $\sq{q}=(\sT,s_L)$, with $s_L$ being the longitudinal component. The remnant quark $q_1$ has momentum $z_1,\vect{p}_{1\perp}$ with respect to $q$ and polarization $\sq{1}$. Then, $q_1$ emits hadron $h$ carrying its light-cone momentum fraction $z_2$ and transverse momentum $\vect{p}_{2\perp}$ with respect to its 3-momentum. The direction of the 3-momentum of $q$ is denoted by $\hat{\vect{z}}$, while that of $q_1$ by $\hat{\vect{z}}'$, as depicted in Fig.~\ref{PLOT_Q_to_Q_H}.

 The momentum components of the second produced hadron $h$ with respect to $q$ are given by 
\al{
z&=z_1 z_2,\\
\vect{p}_\perp& = z_2 \vect{p}_{1\perp} + \vect{p}_{2\perp}.
}

Then we can write the probability density for such a process as
\al
{
\label{EQ_Q_to_2Q}
F^{(2)}_{q\to h}(z, \pp{}; \vect{s} ) &
\\ \non
=\sum_{q_1}&  \hat{f}^{q\to q_1}(z_1, \pe{1};  \sq{q} )  \otimes \hat{f}^{q_1\to h}(z_2, \pe{2}; \sq{q_1} ).
}
 
 The explicit form for the quark-to-quark probability density~\Eq{EQ_PROB_UNPOL} has a Collins-like modulation,
\al
{
\label{EQ_SPLIT_Q_Q}
\hat{f}^{q\to q_1}(z, \pp{}; \vect{s} )&
\\ \non
= \hat{D}^{(q\to q_1)}&(z, \psq{})  + \frac{(\pe{}\times \sT) \cdot \hat{\vect{z}} }{z\mathcal{M}}  \ \hat{H}^{\perp({q\to q_1})}(z,\psq{}),
}
and  the quark-to-unpolarized hadron probability is also given by the familiar form
\al
{
\label{EQ_SPLIT_Q_H}
\hat{f}^{q\to h}(z,\pp{};\vect{s}) &
\\ \non
= \hat{D}^{(q\to h)}&(z, \psq{})  + \frac{(\pe{}\times \sT) \cdot \hat{\vect{z}}}{z m_h}\ \hat{H}^{\perp({q\to h})}(z,\psq{}).
}

Clearly, the unpolarized and Collins terms in these two expressions are related to each other~\cite{Bentz:2016rav}. 

We can use the expressions in Eqs.~(\ref{EQ_EL_ALPH_BETA}),~(\ref{EQ_REMNAT_SPIN}),~(\ref{EQ_SPLIT_Q_Q}), and~(\ref{EQ_SPLIT_Q_H}) to prove that the probability density for $h$ can be written as a sum of two terms
\al
{
F_{q\to h}^{(2)}(z, \pe{};  \sq{q} )  &
\\ \non
= 
D_{q\to h}^{(2)}(z, &\psq{} )  
+ \frac{1}{z m_h} (\pe{}\times \sqt{}) \cdot \hat{\vect{z}}\ H_{q\to h}^{\perp (2) }(z, \psq{}),
}
where $D^{(2)}$ and $H^{\perp(2)}$ correspond to the unpolarized and Collins function for the hadron $h$ at rank 2. These can be expressed in terms of the elementary TMD SF functions 
\begin{widetext}
\al
{
\label{EQ_D_R2}
  {D}^{(2)}_{q\to h}(z,\psq{}) 
  = &
   2 \sum_{q_1}\int_0^1 dz_1 \int_{0}^{1} d z_2 \int d^2\pe{1}  \int d^2\pe{2} 
\times \delta(z -z_1 z_2)\ \delta^2(\pp{} - z_2 \pe{1}- \pe{2}) 
\\\non&
\times \Bigg[
\hat{D}^{q\to q_1} (z, \psq{1}) \ \hat{D}^{q_1 \to h}(z_2, \psq{2})+
 \frac{1}{z \mathcal{M} m_h} (\pe{1} \cdot \pe{2}) \hat{D}_T^{\perp (q\to q_1)}(z_1, \psq{1})  \hat{H}^{\perp (q_1 \to h)}(z_2, \psq{2})
\Bigg],
}
where we sum over all possible intermediate quarks $q_1$.

Then integral expression for the Collins function reads
\al
{
\label{EQ_HP_R2}
  H^{\perp (2)}_{q\to h}&(z,\psq{}) =
 2\frac{z\ m_h}{(\pe{}\times \sT) \cdot \hat{z}} \sum_{q_1}\int_0^1 dz_1 \int_{0}^{1} d z_2 \int d^2\pe{1}  \int d^2\pe{2} 
\times \delta(z -z_1 z_2)\ \delta^2(\pp{} - z_2 \pe{1}- \pe{2}) 
\\\non&
\times \Bigg[
 \frac{1}{z_1\mathcal{M}} (\pe{1} \times \sT)\cdot \hat{\vect{z}}\ \hat{H}^{\perp (q \to q_1)}(z_1, \psq{1})
 \hat{D}^{(q_1 \to h)}(z_2, \psq{2})
\\\non&
+ \frac{1}{z_2 m_h} \Bigg( \pe{2}\times\Bigg\{{\sT \ \hat{H}_T^{(q\to q_1)}(z_1,\psq{1})  +\pe{1} (\pe{1} \cdot \sT) \frac{1}{z_1^2 \mathcal{M}^2} \ \hat{H}_T^{\perp (q\to q_1)}(z_1,\psq{1})}\Bigg\}
\Bigg)
 \cdot \hat{\vect{z}}\ \hat{H}^{\perp (q_1\to h)}(z_2,\psq{2})
\Bigg].
}

\end{widetext}

\section{Models for Elementary Splittings and Positivity Constraints}
\label{SEC_SPLITTINGS}

 The quark-jet framework requires elementary fragmentation functions as input. To calculate the polarization of the intermediate quarks, we need access to all 8 elementary SFs for quark-to-quark transitions, both T-even and T-odd. The corresponding two quark-to-hadron TMD SFs should be related to these to preserve momentum and isospin. In general, we can use any quark model calculations or parametric forms to best reproduce the observables. In this work we use the tree-level  spectator-type calculations of the T-even functions within the NJL effective quark model~\cite{Bentz:2016rav}.  The T-odd SFs  calculated at the same level yield vanishing results~\cite{Meissner:2010cc,Bentz:2016rav}, similar to the case of quark-to-hadron SFs~\cite{Collins:1992kk}. In these models one-loop interference-type cut diagrams are needed to generate nonzero T-odd functions, see e.g.~\cite{Amrath:2005gv, Meissner:2010cc}. To summarize, to date the model calculations of the T-even SFs involve only the tree-level cut diagrams, while those for the T-odd functions involve only the one-loop interference-type diagrams. This yields model elementary TMD SFs that violate the positivity bound of the overall polarized SF, making it impossible to use them for MC simulations.
 
\subsection{Positivity bounds for splitting functions}
\label{SUBSEC_POSITIVITY}  

To demonstrate the violations of the positivity bound by mixed-order calculations for the elementary T-even and T-odd functions we employ the positivity relations derived in Ref.~\cite{Bacchetta:1999kz} to impose constraints on the FFs, which should also hold for model SFs. First let us define the notation for any function $F (z,\psq{})$,
 \al
{
 F^{[1]}  (z,\psq{})  &\equiv \frac{\psq{}}{2 z^2 M^2} F (z,\psq{}) ,
\\
  H(z,\psq{}) &\equiv H_T(z,\psq{}) + H_T^{\perp [1]}(z, \psq{}).
}

 The relations of Ref.~\cite{Bacchetta:1999kz} for the TMD FFs  can then be expressed as
\al
{
\label{EQ_FF_INEQUALITY_H}
 | H | \leq & \frac{1}{2} ( D+ G_L ) \leq D ,
 }
 \al{
  \label{EQ_FF_INEQUALITY_HTP}
  |H_T^{\perp [1]} | \leq &\frac{1}{2} (D - G_L) \leq D ,
}
\al{ 
 \label{EQ_FF_INEQUALITY_HP}
 ( G_T^{[1]} )^2 + ( H^{\perp [1]} )^2 
 &\leq \frac{\psq{}}{4 z^2 M^2} (D + G_L) (D - G_L) 
 \\ \non
&\leq \frac{\psq{}}{4 z^2 M^2}  D^2,
}
\al{
\label{EQ_FF_INEQUALITY_DTP}
( H_L^{\perp [1]} )^2 + ( D_T^{\perp [1]} )^2  &\leq \frac{\psq{}}{4 z^2 M^2} (D + G_L) (D- G_L)
\\ \non
 &\leq \frac{\psq{}}{4 z^2 M^2}  D^2.
}
We can check the validity of relevant relations for the T-even SFs in the spectator model using the explicit expressions shown in Eqs.(\ref{EQ_SPLITTINGS_TEVEN_D})-(\ref{EQ_SPLITTINGS_TEVEN_HTP}). Here we find
\al
{
 | \hat{H} | = \frac{1}{2} ( \hat{D}+ \hat{G}_L ) \leq \hat{D},
}
so the first part of the "Soffer bound" in \Eq{EQ_FF_INEQUALITY_H} is saturated.

Furthermore,
\al
{
 |\hat{H}_T^{\perp [1]} | = \frac{1}{2} (\hat{D} - \hat{G}_L) \leq \hat{D} ,
}
with the first parts of the inequality again being satisfied at the limit (of equality).

We can then easily calculate the expressions
\al
{
 ( \hat{G}_T^{[1]} )^2 =  ( \hat{H}_L^{\perp [1]} )^2  =& \frac{\psq{}}{4 z^2 M^2} (\hat{D} + \hat{G}_L) (\hat{D} - \hat{G}_L) 
 \\ \non
&\leq \frac{\psq{}}{4 z^2 M^2}  \hat{D}^2 ,
}
whichm according to the relations in Eqs.~(\ref{EQ_FF_INEQUALITY_HP})~and~(\ref{EQ_FF_INEQUALITY_DTP}) requires that
\al
{
   \hat{H}^{\perp}(z, \psq{})   &= 0,
   \\
   \hat{D}_T^{\perp} (z, \psq{}) &=0.
}

 In conclusion, in order to satisfy the positivity constraints for the tree-level calculations of T-even SFs in Eqs.~(\ref{EQ_SPLITTINGS_TEVEN_D})-(\ref{EQ_SPLITTINGS_TEVEN_HTP}), the  T-odd functions  $\hat{H}^{\perp}$ and $\hat{D}_T^{\perp}$ should both vanish. Thus, the previous model calculations of these functions violate the positivity constraints. This result does not depend on the details of the regularization of the transverse momentum dependence, etc. 

 This observation is analogous to the case of the TMD PDFs, where the violation of the positivity by the mixed-order model calculations of the T-even and T-odd functions has been known for some time (see Refs.~\cite{Kotzinian:2008fe,Ellis:2008in,Bacchetta:2008af,Pasquini:2011tk}).
 
\section{Monte Carlo Simulations and Results}
\label{SEC_RESULTS}

 The results from Sec.~\ref{SUBSEC_POSITIVITY} indicate that the straightforward leading-order quark model calculations of the elementary TMD FFs violate positivity and cannot be used for MC simulations, most likely because of the mixed-order calculations of the T-even and T-odd ones. Further investigation of this problem, including the next-order calculations of the T-even FFs, is beyond the scope of this work. Thus, here we choose to use ansatz FFs based on the NJL model calculations. First, to accommodate any nonzero T-odd FFs, we slightly increase the unpolarized FF by a constant factor
\al
{
  \hat{D}(z) = 1.1 \ \hat{D}_{tree}(z),
}
where $\hat{D}_{tree}(z)$ is the tree-level result, regularized in the extended Lepage-Brodsky scheme of Ref.~\cite{Matevosyan:2011vj}. For MC simulations we use the numerical values of model parameters from the same article.

For the Collins function we choose a simplistic ansatz, which satisfies the kinematic conditions outlined in Ref.~\cite{Collins:1992kk}
\al
{&
\label{EQ_COLLINS_ANSATZ}
  \frac{p_\perp}{z M}\frac{\hat{H}^{\perp (q\to h)}(z,\psq{})}{\hat{D}^{(q\to h)}(z,\psq{}) } = 0.4\ \frac{2\ p_{\perp} M_Q }{\psq{} + M_Q^2} ,
}
where $M_Q$ is the mass of the remnant quark, and the coefficient $0.4$ is chosen to satisfy the positivity constraints in Eqs.~(\ref{EQ_FF_INEQUALITY_H})-(\ref{EQ_FF_INEQUALITY_DTP}). Additionally, we use the model relations
\al
{
  \hat{D}^{(q\to h)}(z,\psq{})  &= \hat{D}^{(q\to q_1)}(1-z,\psq{}) ,
  \\
  \hat{H}^{\perp (q\to h)}(z,\psq{})  &= - \hat{H}^{\perp (q\to q_1)}(1-z,\psq{}),
  \\
  \hat{H}^{\perp (q\to q_1)}(z,\psq{}) &= - \hat{D}^{\perp (q\to q_1)}_T(z,\psq{}),
}
where all the elementary splittings are assumed to be normalized, as described in Ref.~\cite{Matevosyan:2011vj}.

 In this work we present only $\psq{}$-integrated quantities for brevity. Then the relevant number density becomes
\al
{
\label{EQ_POL_FF_COL}
   D_{h/q^\uparrow}(z, \vf) &= \frac{1}{2\pi} \Big[   {D}_{q\to h}(z) -2  {H}^{\perp (1/2)}_{q\to h}(z)  s_T \sin(\vf_C)  \Big],
}
where
\al{
\label{EQ_UNPOL_INTEGR}
    {D}_{q\to h}(z) &=\pi  \int _0^\infty d \psq{}  D (z, \psq{}),
   \\
   \label{EQ_COL_ONEHALF}
 H^{\perp (1/2)}(z) &=  \pi \int _0^\infty d \psq{} \frac{p_\perp}{2 z m_h} H^\perp (z, \psq{}).
}

 We can extract the collinear unpolarized FF ${D}_{q\to h}(z)$ and the $1/2$ moment of the Collins function $ {H}^{\perp (1/2)}_{q\to h}(z)$ by fitting the relation~(\ref{EQ_POL_FF_COL}) with a functional form linear in $\sin(\vf_C)$,
\al
{
\label{EQ_SIN_LIN}
 L(z,\vf_C) = c_0(z) - c_1(z) \sin(\vf_C).
}
We can easily access the full TMD FFs of \Eq{EQ_Dqh_SIN} using the same method from unintegrated number densities we compute, but here we focus on presenting the overall MC framework and reserve presenting full TMD results for future work.

For simplicity, we  calculate only the FFs of the $u$ quark to pions in this work. In our simulations we use $100$ points to discretize $z$ and $\psq{}$, and $600$ points for the azimuthal angles. To achieve high precision results needed for validating various aspects of the MC framework, for each of the computations presented here we simulated at least $10^{11}$ hadronization chains by running the MC software in parallel on a small computer farm.

\subsection{Quark-to-hadron FFs for rank-2 hadrons}
\label{SUBSEC_VERIFY_2STEP}

\begin{figure}[htb]
\centering 
\subfigure[]
 {
\includegraphics[width=\ImL]{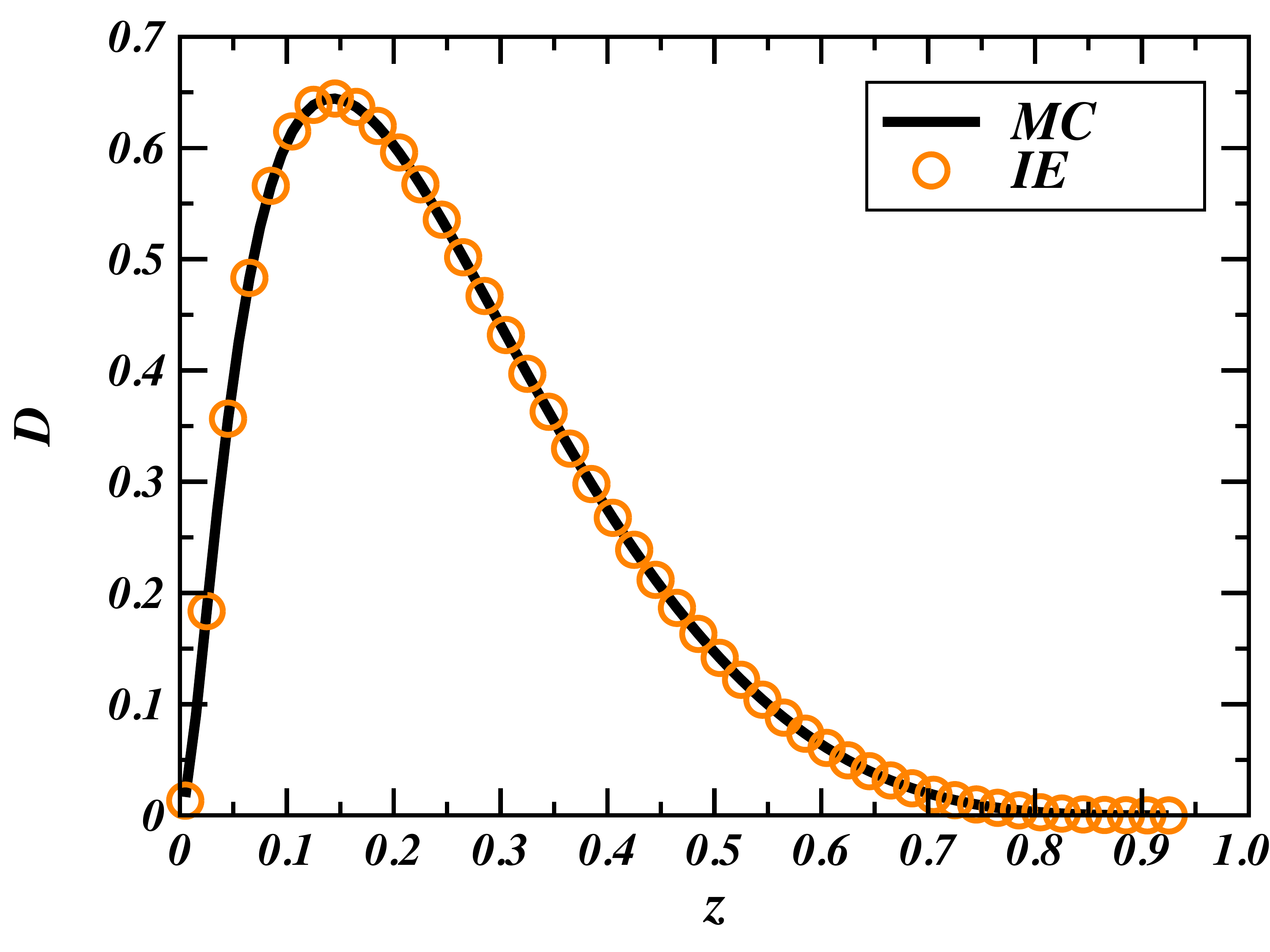}
}
\\ \GapSubf
\subfigure[]
 {
\includegraphics[width=\ImL]{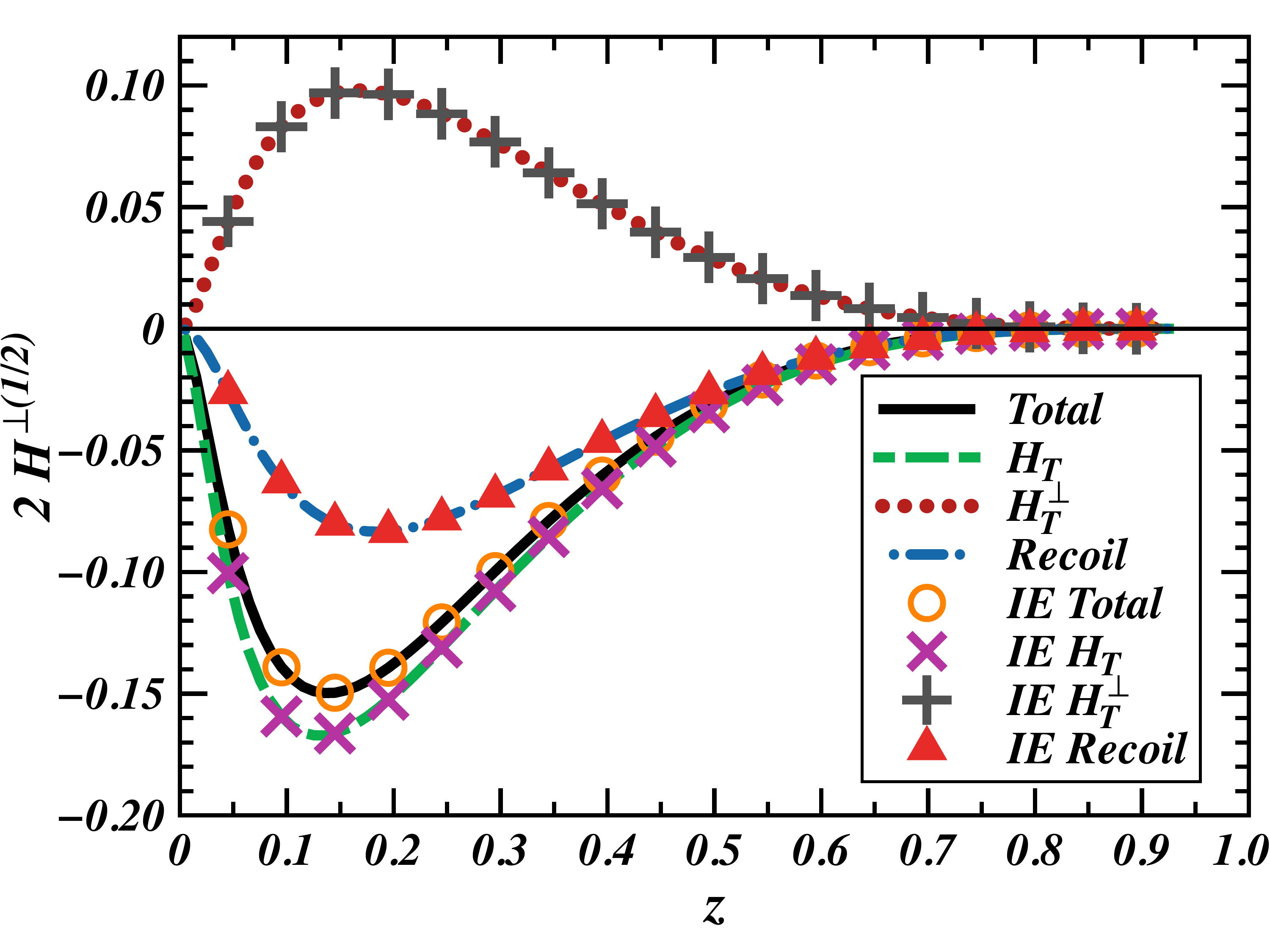}
}
\\ \GapSubf
\subfigure[] {
\includegraphics[width=\ImL]{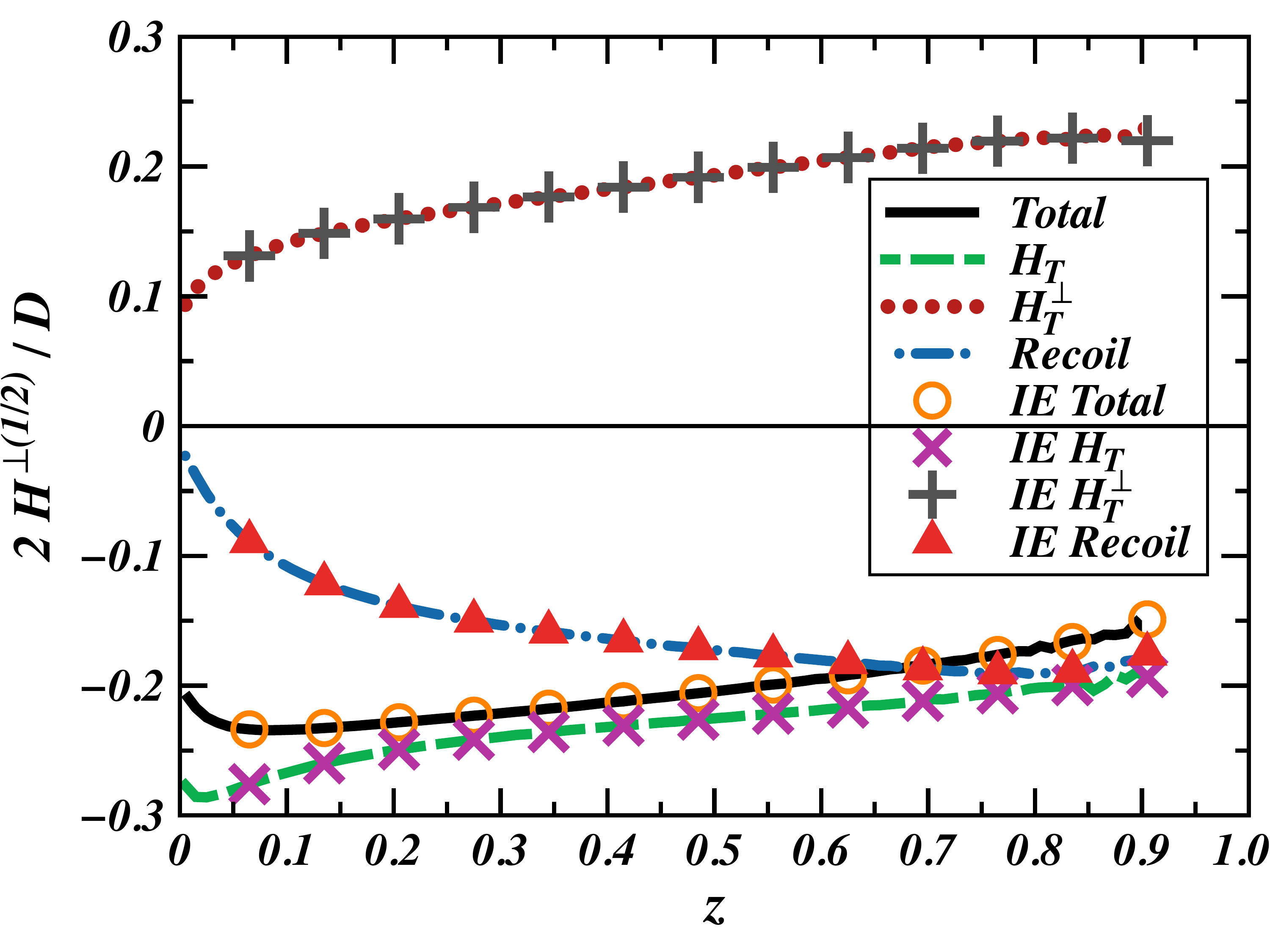}
}
\GapCapt
\caption{Comparison of results of  $D$ (a), $H^{\perp (1/2)}$ (b), and  $H^{\perp (1/2)}/D$ (c) of integral expressions (IE) and MC results  for $\pi^+$ produced at rank 2 in the quark-jet picture. The plots in (b) and (c) show the different contributions to the Collins function at rank 2 shown in~\Eq{EQ_HP_R2}.}
\label{FIG_H12_R2}
\end{figure}

First, we verify the MC framework by comparing the results for the unpolarized and Collins functions of the rank-2 $\pi^+$ produced by the $u$ quark with those obtained via numerical integration of relations~(\ref{EQ_D_R2}) and (\ref{EQ_HP_R2}). The corresponding plots, which also  compare the contributions of the three terms in~(\ref{EQ_HP_R2}), are shown in Fig.~\ref{FIG_H12_R2}. Here, the label "RECOIL" refers to the term involving $\hat{H}^{\perp (q\to q_1)} \otimes \hat{D}^{(q_1\to h)}$, while $\hat{H}_T$ and $\hat{H}_T^\perp$ refer to the  terms where the corresponding functions are convoluted with $\hat{H}^{\perp (q_1\to h)}$. The plots show a perfect agreement between the MC results and those from Eqs.~(\ref{EQ_D_R2}) and (\ref{EQ_HP_R2}). Moreover, we see that the recoil term is almost canceled by the one involving $\hat{H}_T^\perp$. Further, we see that the recoil term in the analyzing power $2 H^{\perp (1/2)} / D$ vanishes as $z\to 0$, while the other two remain nonzero.

\subsection{Higher order modulations in a naive model}
\label{SUBSEC_SPIN_FLIP}

\begin{figure}[tb]
\centering 
\includegraphics[width=\ImL]{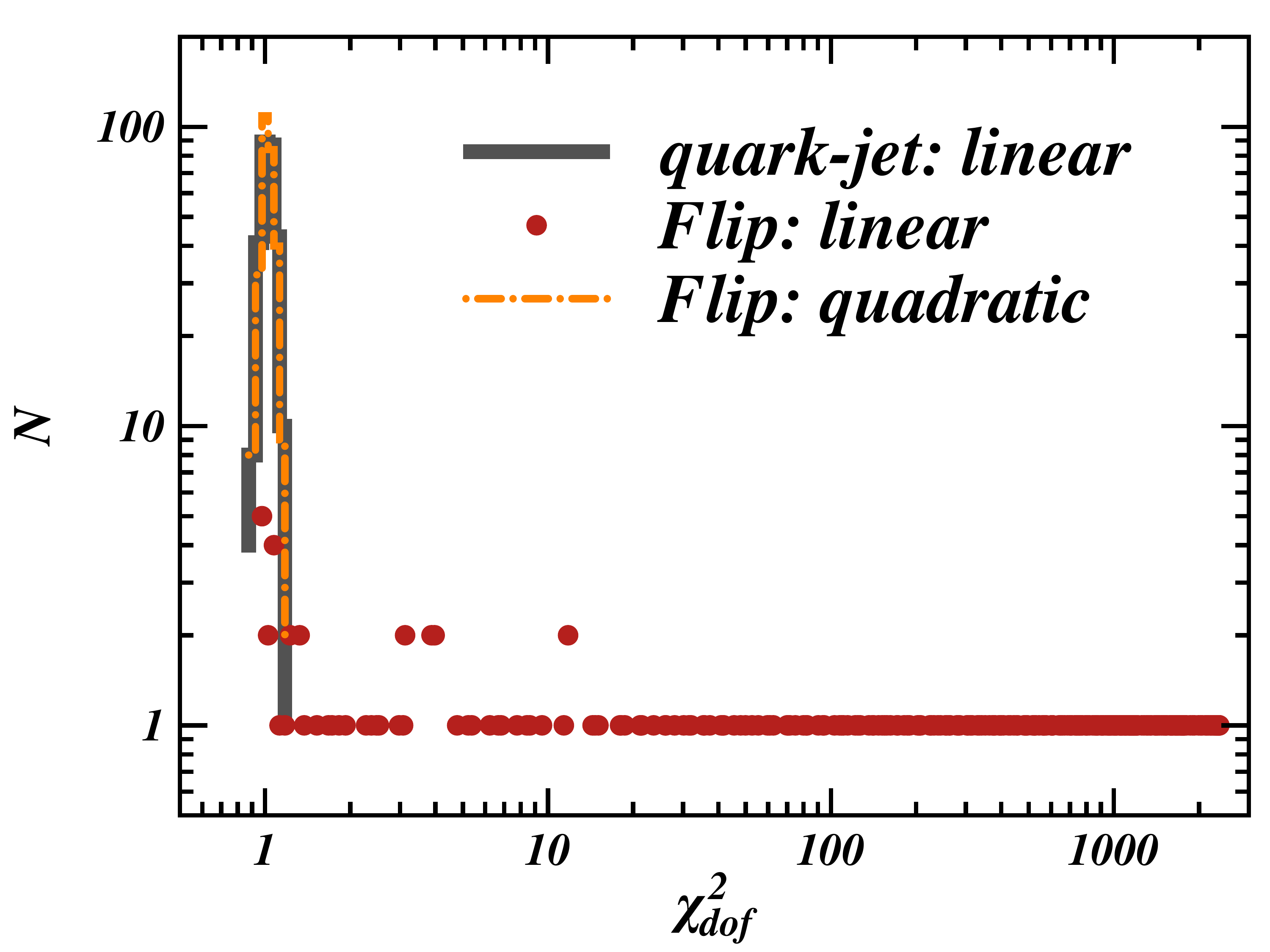}
\GapCapt
\caption{Histogram of the values of $\chi^2_{\rm dof}$ for fits of all polarized fragmentation functions of the $u$ quark to rank-2 pions, fitted with linear and quadratic polynomials in $\sin(\vf_C)$ of Eqs.~(\ref{EQ_SIN_LIN}) and (\ref{EQ_SIN_QUAD}) for MC simulations of rank-2  hadrons. The label "Flip" denotes the simulations where the transverse polarization of the quark is simply flipped after each hadron emission step.}
\label{PLOT_ChiSq_Ranks}
\end{figure}

 One of the main motivations for this work was to extend the quark-jet formalism to include the quark polarization propagation in a self-consistent manner that does not induce unphysical higher-order $\sin(\vf_C)$ modulations, as discussed in the Introduction. If present, these modulations should appear for hadrons at rank 2 and higher. To test our model, we calculated the values of $\chi^2_{\rm dof}$ for fits to the polarized FF in (\ref{EQ_POL_FF_COL}) for every value of z and all pions at rank 2, about $300$ fits in total (we skip the few $z$ bins with a very small number of events for a given hadron that would yield very large statistical uncertainties). The results are shown in Fig.~\ref{PLOT_ChiSq_Ranks}, where the $\chi^2_{dof}$ values are sharply peaked around $1$, with a maximum value not exceeding $1.5$. Also in this figure, the histograms labeled "Flip" are the results for the naive model, where the transverse component of the fragmenting quark's polarization is simply flipped (the azimuthal angle is increased by $\pi$) after every emission step, while its modulus and the longitudinal components are unchanged.  We readily see that the linear form in $\sin(\vf_C)$ of \Eq{EQ_SIN_LIN}  fails to provide reasonable fits to the MC results, while the quadratic form
\al
{
\label{EQ_SIN_QUAD}
 Q(z,\vf_C) = c_0(z) - c_1(z) \sin(\vf_C)- c_2(z) \sin^2(\vf_C),
}
 fits perfectly. This once again demonstrates the presence of the  higher-order modulations in models where the polarization transfer of the quark to its remnant is not correctly described. We also explicitly checked, that the linear function produces high quality fits to hadrons of all the ranks in our simulations with the self-consistent formalism.

\subsection{Results for the FFs}
\label{SUBSEC_MAIN_RESULTS}

Here we present the results for the model calculations with $N_L=10$ hadron emissions. Also, to demonstrate the flexibility of our model, we include results where the input model FFs have been multiplied by $(1-z)^4$. This modified model mimics the effects of QCD evolution, where the unpolarized FF is peaked at low values of $z$, unlike in our unevolved model~\cite{Matevosyan:2010hh}. 

\begin{figure}[tb]
\centering 
\subfigure[] {
\includegraphics[width=\ImL]{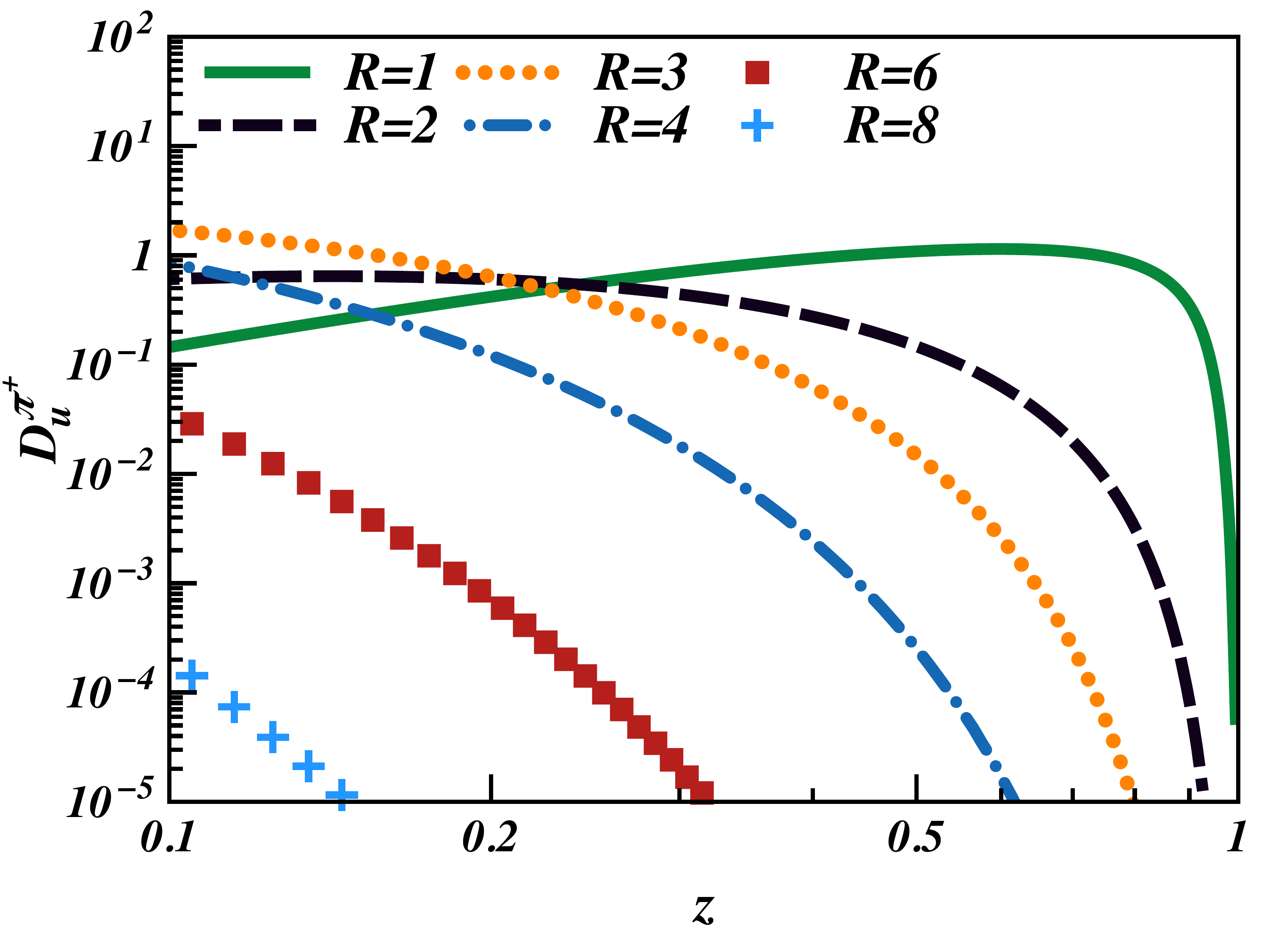}
}
\\ \GapSubf 
\subfigure[] {
\includegraphics[width=\ImL]{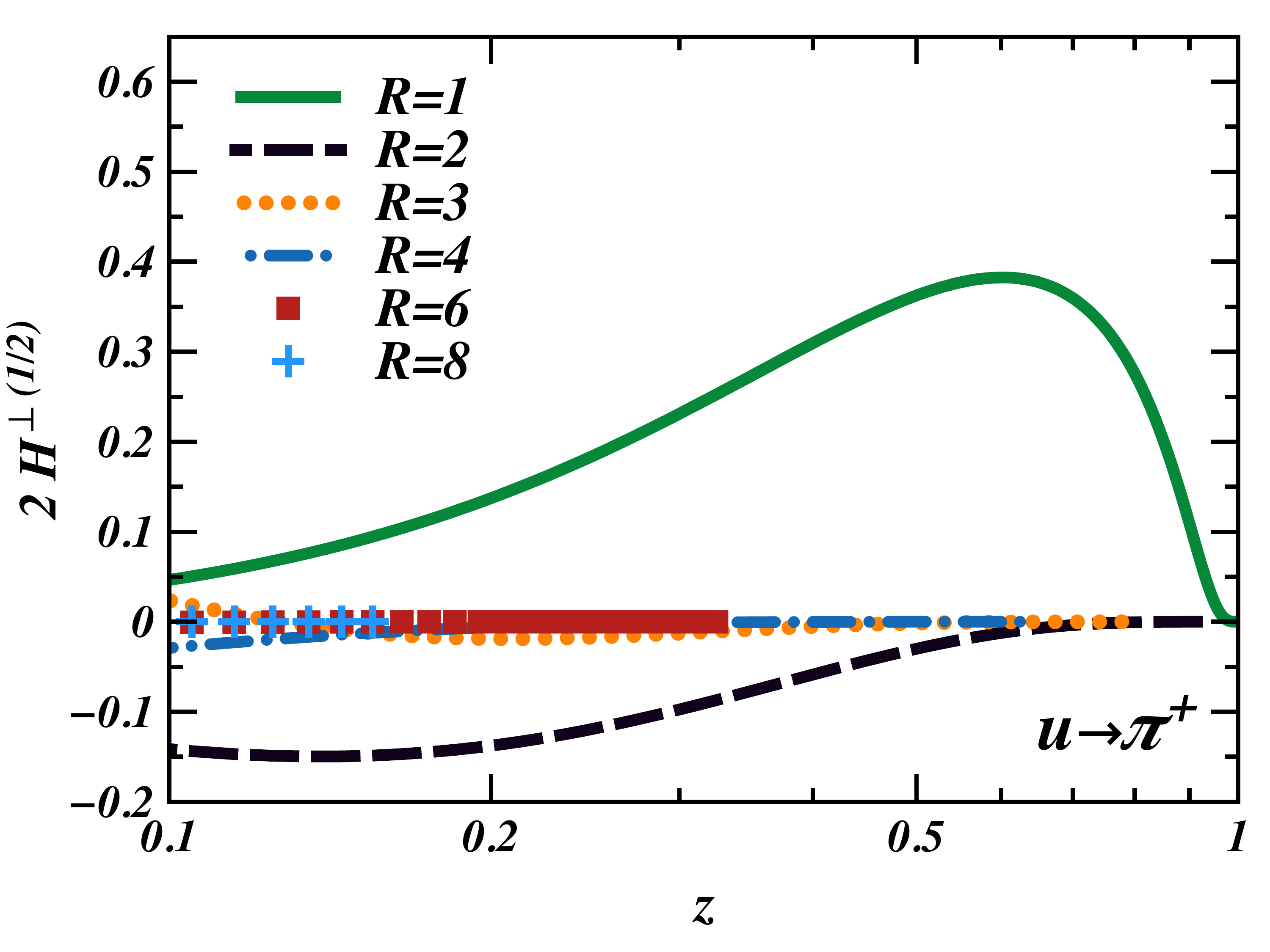}
}
\GapCapt
\caption{Fitted values of $D$  (a)  and $2 H^{\perp (1/2)}$ (b) for $u\to\pi^+$ as a function of $z$for hadrons at different ranks from Monte Carlo simulations using model SFs.}
\label{PLOT_FRAG_RANK_X}
\end{figure}

\begin{figure}[tb]
\centering 
\subfigure[] {
\includegraphics[width=\ImL]{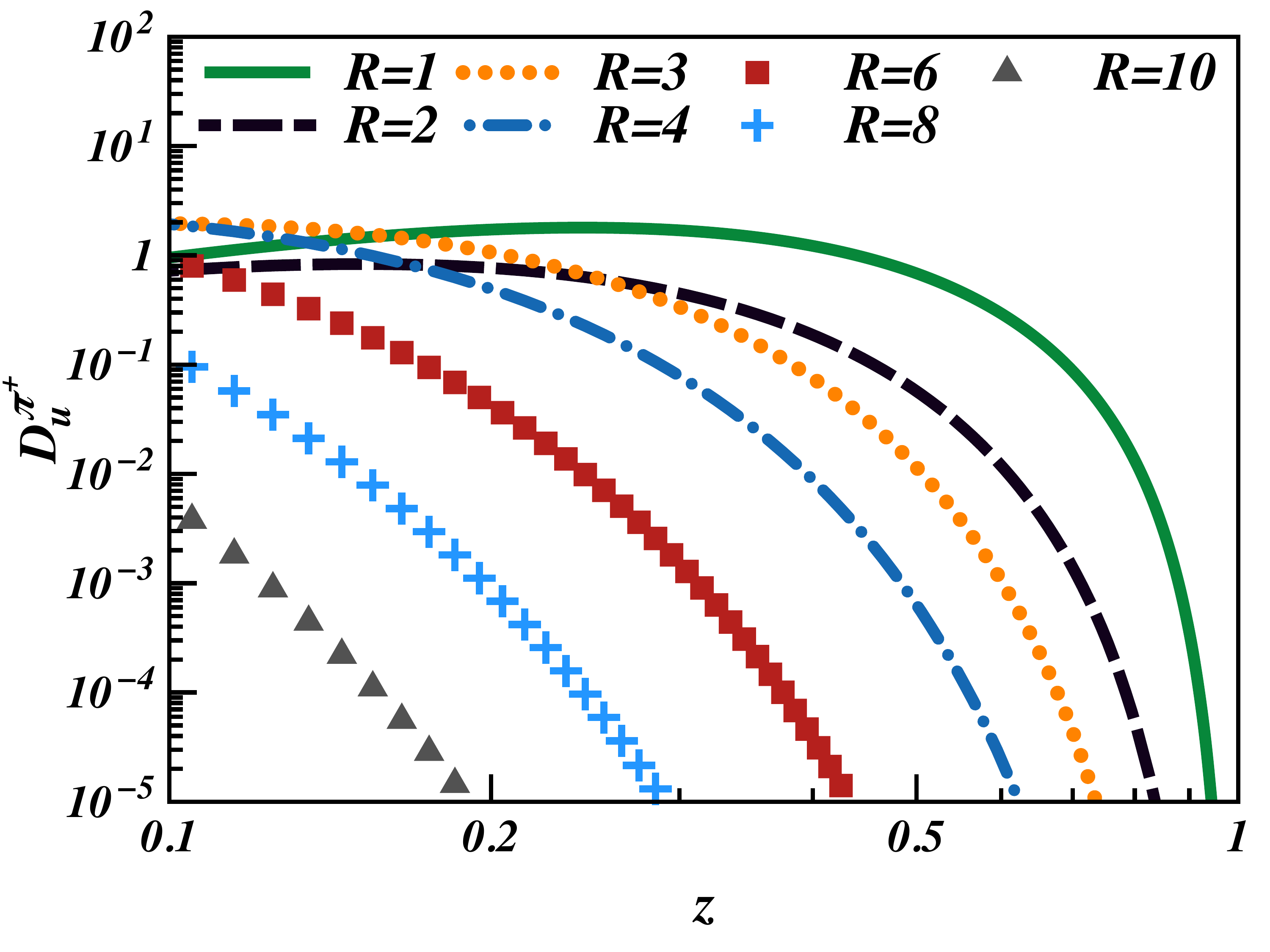}
}
\\ \GapSubf 
\subfigure[] {
\includegraphics[width=\ImL]{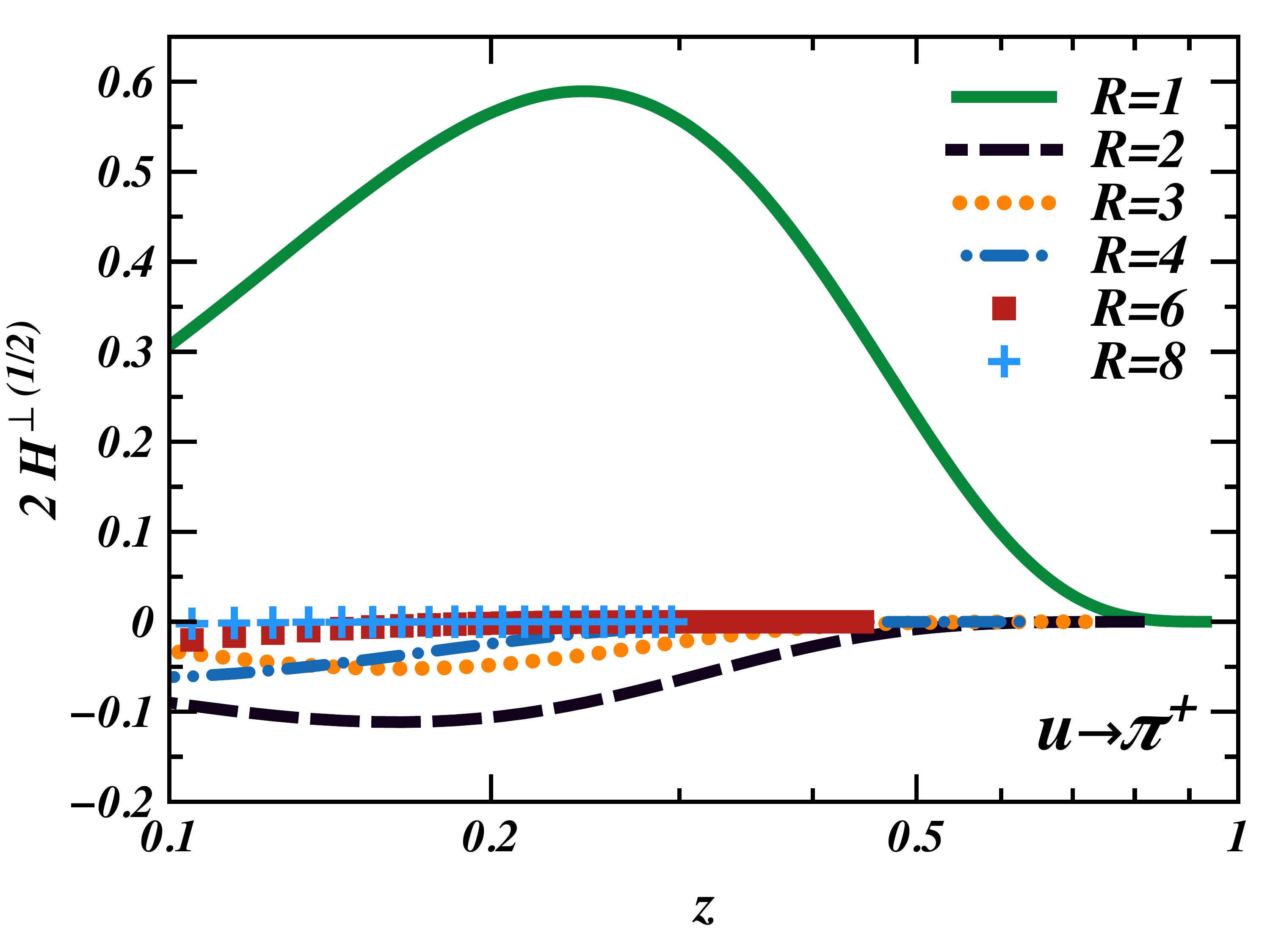}
}
\GapCapt
\caption{Fitted values of $D$  (a)  and $2 H^{\perp (1/2)}$ (b) for $u\to\pi^+$ as a function of $z$ for hadrons at different ranks from Monte Carlo simulations using model SFs modified by a factor $(1-z)^4$ .
}
\label{PLOT_FRAG_RANK_X_MIZ_4}
\end{figure}

 The plots in Figs.~\ref{PLOT_FRAG_RANK_X} and \ref{PLOT_FRAG_RANK_X_MIZ_4} show the unpolarized and Collins terms of the $\pi^+$ produced at a given rank by an initial $u$ quark in the two models. It is clear that the original model rapidly converges with respect to the rank of the hadron for any reasonably large value of $z$. On the other hand,  the modified model converges slower due to the skewed input unpolarized splitting favoring small $z$ in each hadron emission step.

%
%
%

\begin{figure}[tb]
\centering 
\subfigure[] {
\includegraphics[width=\ImM]{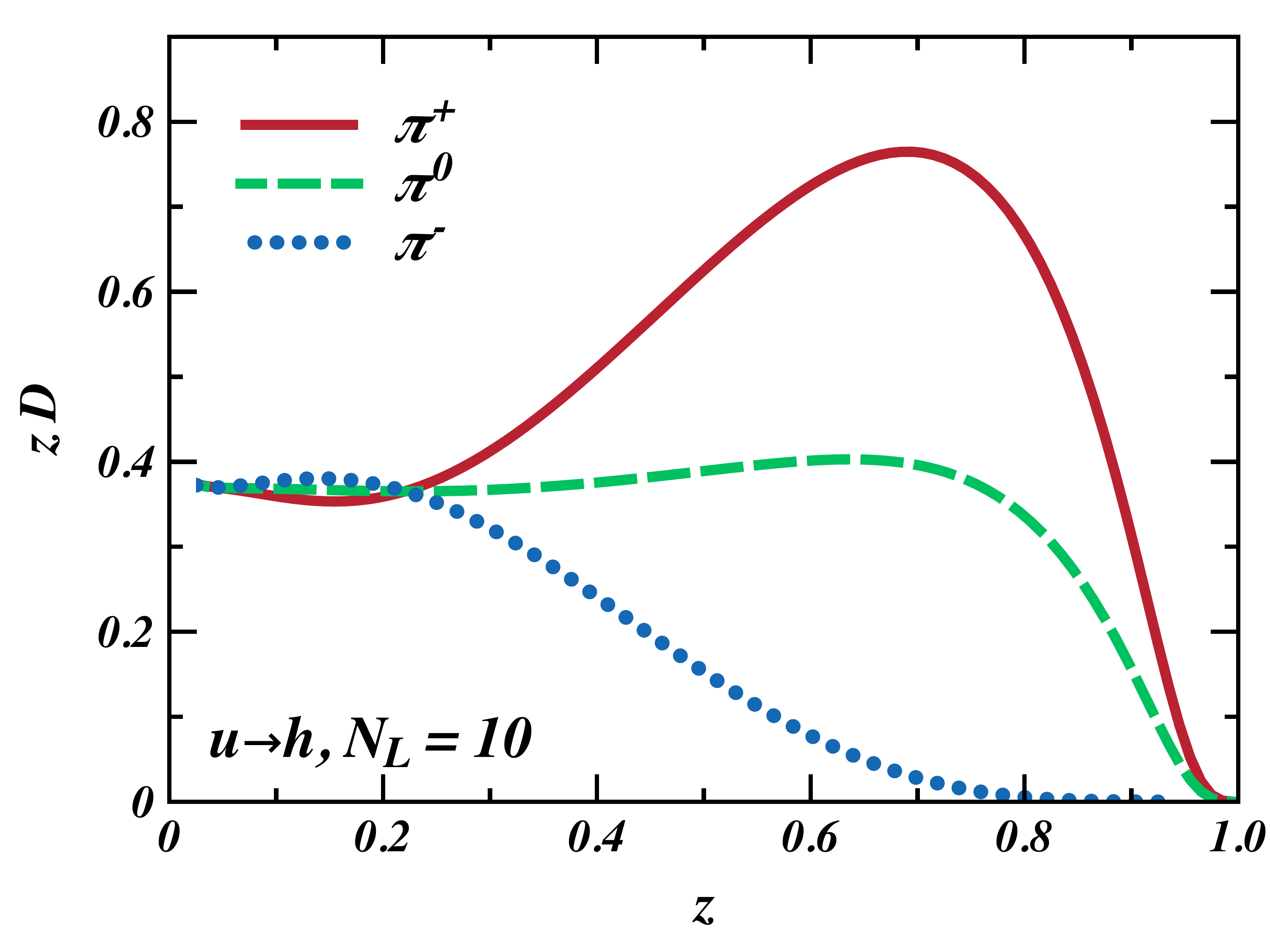}
}
\\ \GapSubf
\subfigure[] {
\includegraphics[width=\ImM]{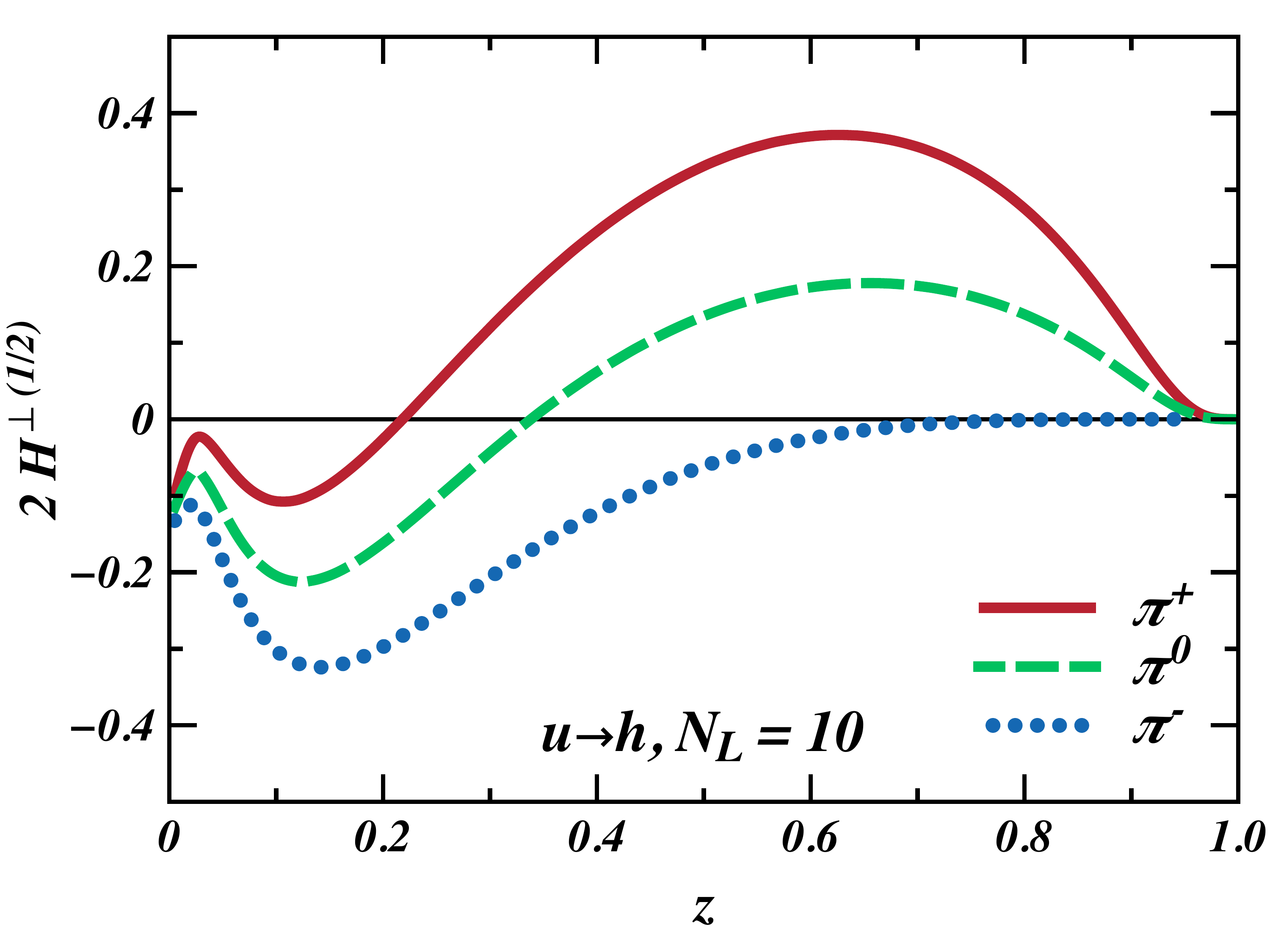}
}
\\ \GapSubf
\subfigure[] {
\includegraphics[width=\ImM]{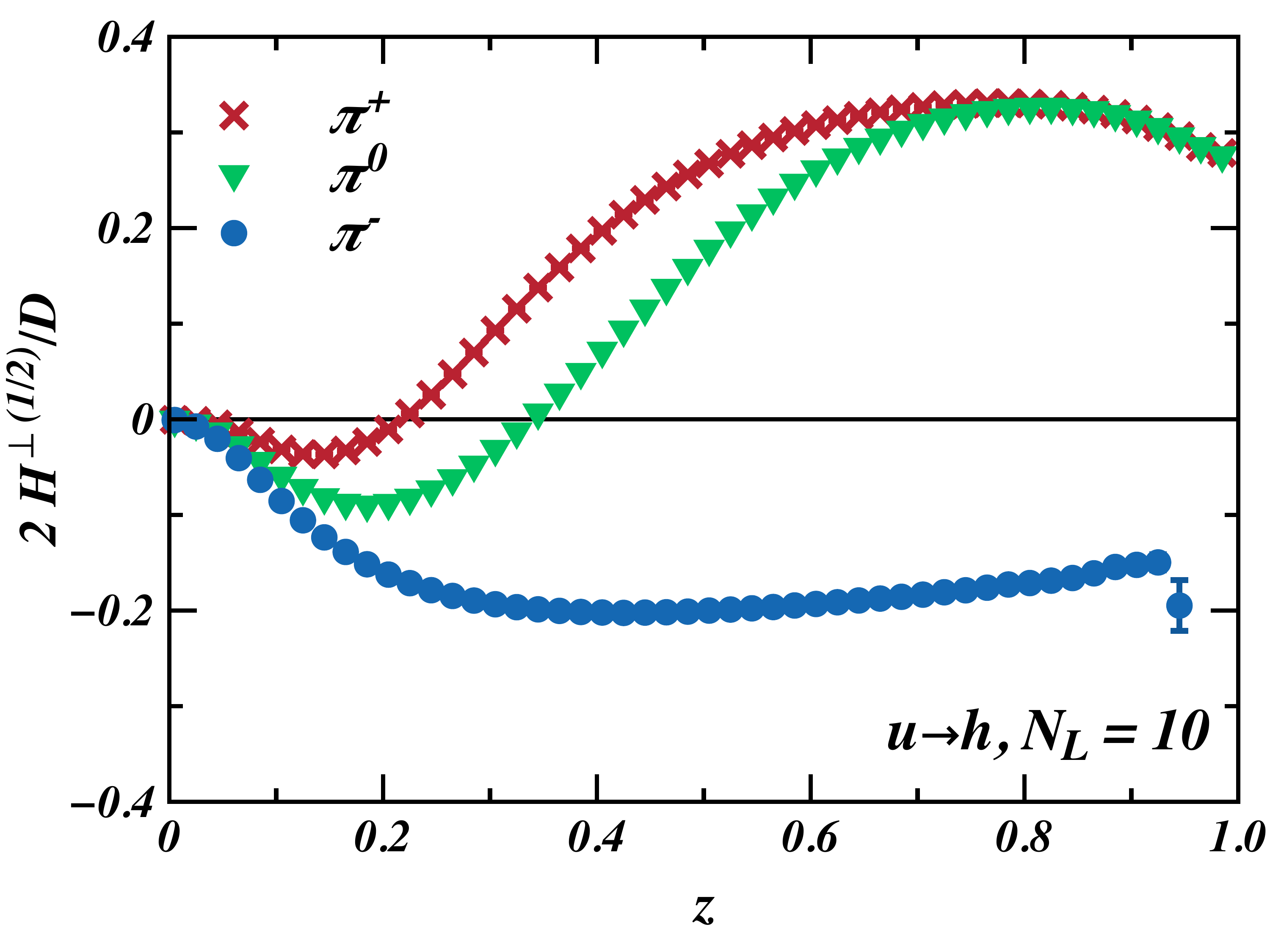}
}
\GapCapt
\caption{Fitted values of $zD$ (a), $2 H^{\perp (1/2)}$ (b),  and  $2H^{\perp (1/2)} / D$ (c) as a function of $z$~from Monte Carlo simulations for $u \to \pi$, with $N_L=10$ emitted hadrons.
}
\label{PLOT_FRAG_PI}
\end{figure}

\begin{figure}[tb]
\centering 
\subfigure[] {
\includegraphics[width=\ImM]{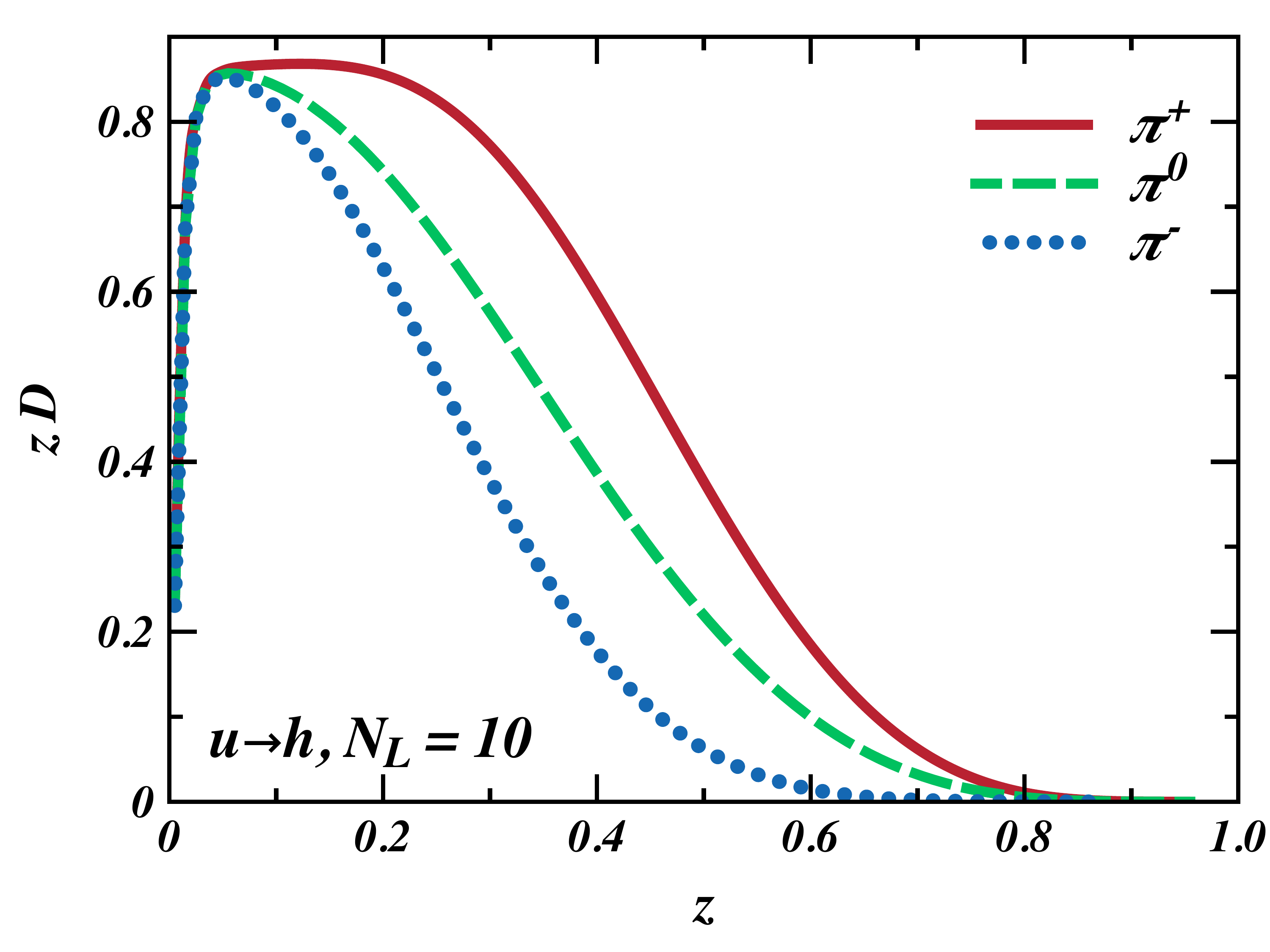}
}
\\ \GapSubf 
\subfigure[] {
\includegraphics[width=\ImM]{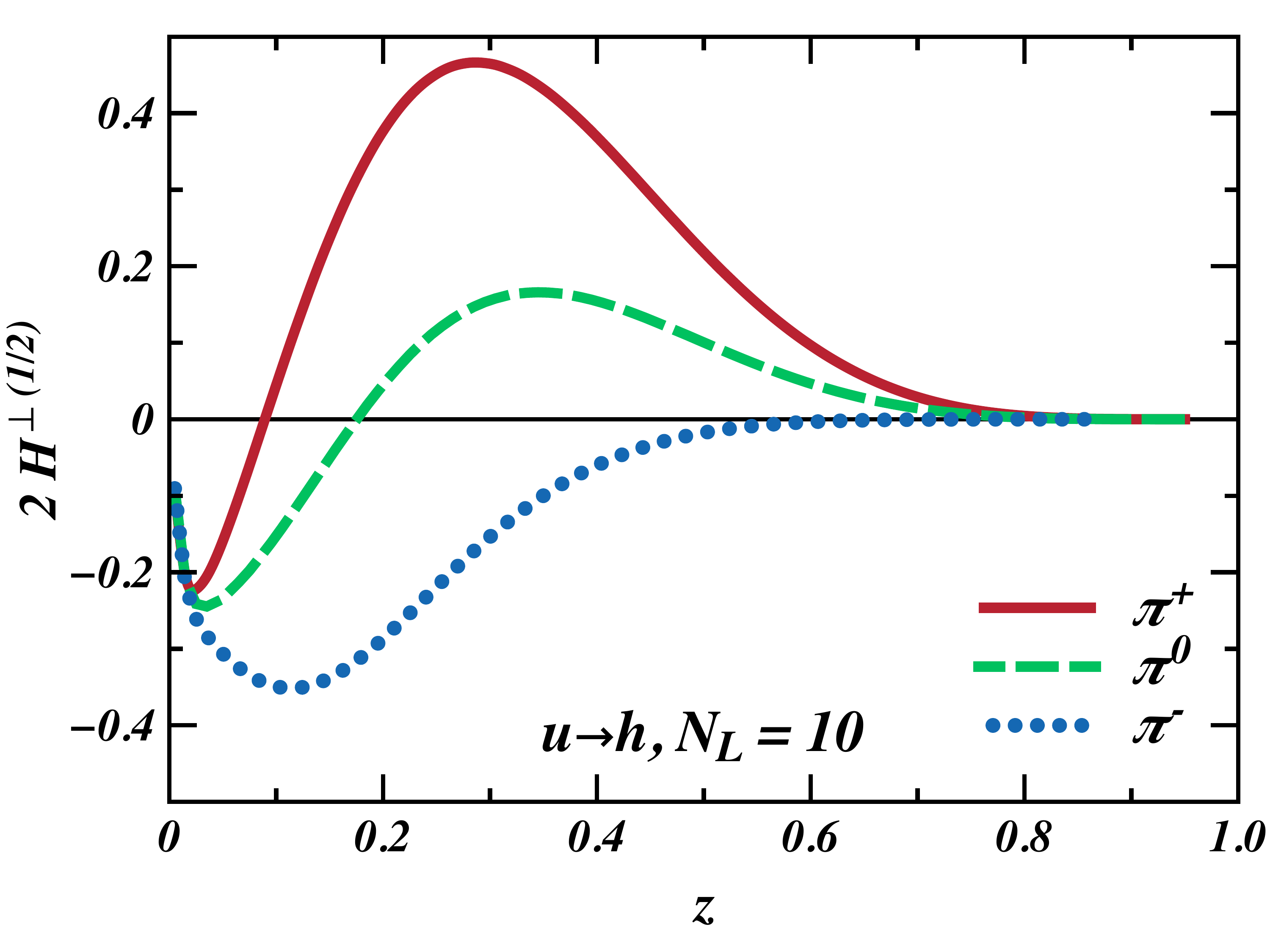}
}
\\ \GapSubf
\subfigure[] {
\includegraphics[width=\ImM]{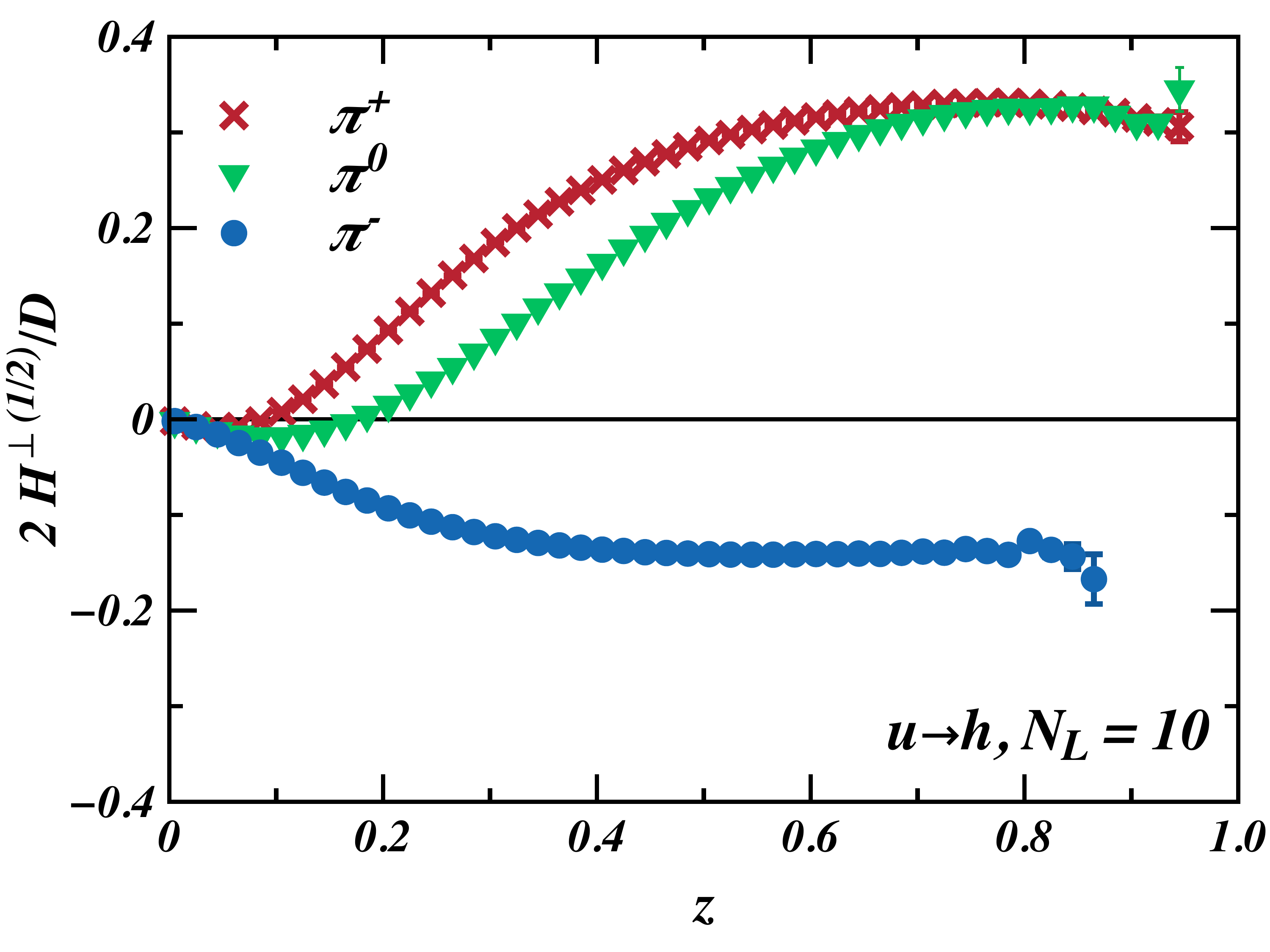}
}
\GapCapt
\caption{The analogous plots to those in Fig.~\ref{PLOT_FRAG_PI} for the modified model.}
\label{PLOT_FRAG_PI_MIZ_4}
\end{figure}

The plots in Figs.~\ref{PLOT_FRAG_PI} and \ref{PLOT_FRAG_PI_MIZ_4} show the unpolarized terms, the Collins terms,  and the analyzing powers of the pions produced by an initial $u$ quark in the two models. We used a large number of produced hadrons, $N_L=10$, in each hadronization chain to ensure complete saturation of the results up to very small $z$ for both models. While the results for both calculations share common features of opposite sign for the favored and unfavored Collins functions at large $z$, the detailed behavior over mid to low values of $z$ can be clearly tuned to best reproduce the data. For example, the results by COMPASS, STAR and BELLE indicate significant asymmetries at $z=0.2$, with opposite signs for the favored and unfavored FFs, which seem to best suit the scenario in the modified model.

%
%

\section{Conclusions}
\label{SEC_CONCLUSIONS}

  The accurate description of the polarized quark hadronization process remains one of the most challenging aspects in the phenomenological description of deep inelastic scattering processes.  For example, the treatment of the quark polarization and the corresponding correlations are, to date, not included in any of the well-known event generators, such as PYTHIA~\cite{Sjostrand:2007gs}, HERWIG~\cite{Bahr:2008pv}, and SHERPA~\cite{Gleisberg:2008ta}.  In this work we presented  Monte Carlo implementation of the extended quark-jet hadronization framework~\cite{Bentz:2016rav}, aimed at calculating various spin-dependent observables such as the Collins FFs. In Sec.~\ref{SUBSEC_INT_QUARK_SPIN} we presented the theoretical framework for the iterative description of the quark-to-quark fragmentation process based on the spin density matrix formalism, and the calculation of the transition probability and the final quark's polarization in terms of the 8 elementary TMD SFs. In Sec.~\ref{SUBSEC_MC_APPROACH} we briefly described the procedure for extracting the unpolarized and Collins FFs of the produced hadrons, similar to our earlier work, and in Sec.~\ref{SUBSEC_TWO_STEP} we presented explicit expressions for the unpolarized and Collins FFs of hadrons produced at rank~2.
  
   One of the important inputs to the quark-jet framework are the 8 elementary quark-to-quark TMD SFs that can be either calculated in effective quark models or parametrized. There are, however, stringent constraints set on these SFs to guarantee the positivity of the total polarized quark fragmentation probability density, derived in Ref.~\cite{Bacchetta:1999kz}. At the same time, the current models describing quark-to-quark and quark-to-hadron fragmentations use the so-called spectator approximation.  Here, the T-even SFs are calculated at the tree level, while the nonvanishing contributions to T-odd functions first appear at one loop. In Sec.~\ref{SUBSEC_POSITIVITY} we prove that the spectator-type quark model calculations of the 6 T-even TMD SFs do satisfy these constraints only if the two T-odd SFs, the Collins and "polarizing" functions, vanish. In this work we circumvented this problem by introducing an ansatz for the TMD SFs, based on NJL model calculations of the T-even functions. 
   
   We presented the results of our MC simulations in Sec.~\ref{SEC_RESULTS} for up-quark fragmentation into pions. First, in Sec.~\ref{SUBSEC_VERIFY_2STEP} we used the explicit results for the rank-2 hadrons of Sec.~\ref{SUBSEC_TWO_STEP} to validate the MC method for calculating both unpolarized and Collins FFs. The plots in Fig.~\ref{FIG_H12_R2} demonstrated that we could precisely reproduce the MC results by calculating the multidimensional integrals numerically. Another important test for these MC simulations is that the resulting polarized FFs  should only depend linearly on the initial quark's polarization and thus also on $\sin(\vf_C)$. In Sec.~\ref{SUBSEC_SPIN_FLIP} we  demonstrated that the linear form of~\Eq{EQ_SIN_LIN} perfectly fits the results of MC simulations, as shown in Fig.~\ref{PLOT_ChiSq_Ranks}. Further, we once again demonstrated in the same figure that, the naive model with transverse polarization flip after each hadron emission, generates unphysical quadratic dependence on the polarization of the initial quark already for the rank-2 hadrons.

 Finally, the full model results were presented in Sec.~\ref{SUBSEC_MAIN_RESULTS}. Here, in addition to the model ansatz for the TMD FFs described in the beginning of Sec.~\ref{SEC_RESULTS}, we also introduced a second form for these functions to demonstrate the flexibility of our MC framework. The plots in Figs.~\ref{PLOT_FRAG_PI} and~\ref{PLOT_FRAG_PI_MIZ_4} showed the corresponding results for the unpolarized and Collins FFs of an up quark to pions, as well as the analyzing powers. The analyzing powers demonstrated the distinctive features of the quark-jet framework: opposite sign for the large $z$ values for  favored and unfavored channels. The results for the favored channel then fall off in magnitude more rapidly than the unfavored ones with decreasing $z$, and they cross the zero at some small $z$. It is also interesting to note that the shapes of the analyzing powers and the zero crossover points for the favored ones drastically depend on the forms of the input splitting functions. For example, the second model ansatz was constructed to mimic the effect of QCD evolution on the unpolarized FFs by skewing the corresponding functions towards the small $z$ region. This yielded roughly equal in magnitude and opposite in sign favored and unfavored analyzing powers for $0.15 \leq z \leq 0.3$, as shown in Fig.~\ref{PLOT_FRAG_PI_MIZ_4}. For example, this behavior looks strikingly similar to the recent results by the STAR Collaboration~\cite{Adkins:2016uxv}, where our original model fails to describe the shape of the experimental results in $z\approx 0.2$. 
 
 Of course,  the experimentally measured single spin asymmetries (SSAs) are ratios of convolutions of PDFs and FFs, and it is naive to directly compare analyzing powers of fragmentation functions to them. A more meaningful comparison to the phenomenological fits of the Collins function to the data of Refs.~\cite{Kang:2015msa,Anselmino:2015sxa}  indeed shows that there are significant discrepancies. Most notably, the zero crossing of the favored Collins function in our model at small $z$ is absent in the parametrizations. There are several reasons for such differences. First, the current functional forms used in Refs.~\cite{Kang:2015msa},~\cite{Anselmino:2015sxa} to fit the Collins FFs do not allow a sign change by construction since the present sizable experimental uncertainties do not allow one to discriminate any possible nodes of the Collins function at small values of $z$. Moreover, the SIDIS experimental measurements to date that are used in the phenomenological fits are  presented for $z\geq 0.2$, leaving the $z <0.2$ behavior of the fitting functions unconstrained. Finally, the measurements are done at substantially larger $Q^2$ than that assumed in the NJL model; thus, we expect the zero-crossing point of the favored FF entering the measured SSA to be pushed to even smaller $z$ by QCD evolution. Similarly,  the scale $Q_0^2=2.4~\Gs$ used to plot the final results in both Refs.~\cite{Kang:2015msa,Anselmino:2015sxa} is significantly  larger than that typically assigned to input elementary FFs calculated in the NJL model ($Q_{NJL}^2\approx 0.2~\Gs$). The "evolution-mimicking" ansatz still should not be expected to give results that can be directly compared with the data at much higher scales, as can be judged by the unpolarized FFs in Figs.~\ref{PLOT_FRAG_PI}(a) and \ref{PLOT_FRAG_PI_MIZ_4}(a). Instead, it is used to demonstrate the flexibility of the framework. In our model we still do not include the vector meson production and strong decays, which should also have a significant influence, especially in the small-$z$ region~\cite{Matevosyan:2011ey}.  Nevertheless, this work was aimed at numerically validating the quark-jet framework approach and to demonstrate the flexibility of the framework, with a perspective that, in the future, a very good description of the experimental data could be achieved by employing flexible functional forms for the input TMD FFs and adjusting the parameters.
 
  The future developments of this model, such as the inclusion of the strange quarks and kaons, as well as the vector meson production and strong decays, will allow one to precisely describe a large range of phenomena that involve polarized quark hadronization. The computation of various polarized dihadron FFs will provide an improved set of predictions compared to our previous work~\cite{Matevosyan:2013eia} with a simplistic model. Further work on the model calculations of the input TMD FFs would give more predictive power to the framework. At the same time, the polarization transfer mechanism used in this work can be readily adapted into the well-known MC event generators such as PYTHIA~\cite{Sjostrand:2007gs}, with parametric forms for the input functions that can be tuned to best reproduce various experimental data.

\section*{Acknowledgements}

 The work of H.H.M. and A.W.T. was supported by the Australian Research Council through the ARC Centre of Excellence for Particle Physics at the Terascale (CE110001104), and by an ARC Australian Laureate Fellowship FL0992247 and Discovery Project No. DP151103101, as well as by the University of Adelaide. A.K. was supported by A.I. Alikhanyan National Science Laboratory (YerPhI) Foundation, Yerevan, Armenia.

\appendix

\section{T-even splitting function from Spectator Model}
\label{SEC_APP_SPLIT}

In spectator-type models, the leading-order elementary T-even TMD SFs are calculated using the cut diagram by evaluating the traces of the quark-quark correlator~\cite{Meissner:2010cc,Bentz:2016rav}. The resulting TMD SFs are shown in Eqs.~(\ref{EQ_SPLITTINGS_TEVEN_D})-(\ref{EQ_SPLITTINGS_TEVEN_HTP}), where we list the "bare" results without any regularization scheme or form factors for the divergent transverse momentum dependence. It is important to note, that any analytic regularization scheme should not affect the inequalities in Sec.~\ref{SUBSEC_POSITIVITY} that follow purely from the relations between the numerators of the splitting functions in~Eqs.(\ref{EQ_SPLITTINGS_TEVEN_D})-(\ref{EQ_SPLITTINGS_TEVEN_HTP}),
\al
{
\label{EQ_SPLITTINGS_TEVEN_D}
\hat{D}(z,\psq{})\ 
=\ & C \Big[ \psq{} + (1-z)^2  M^2\Big],
\\
\label{EQ_SPLITTINGS_TEVEN_GL}
\hat{G}_L(z, \psq{}) 
=\ & C \Big[ -\psq{} + (1-z)^2  M^2\Big],
\\\label{EQ_SPLITTINGS_TEVEN_GT}
\hat{G}_T(z, \psq{})\ 
=\ & C \Big[2 z (1-z) M^2 \Big] ,
}
\al
{
\non \\ \label{EQ_SPLITTINGS_TEVEN_HT}
\hat{H}_T(z,\psq{})\ 
=\ & -\hat{D}(z,\psq{}),
\\\label{EQ_SPLITTINGS_TEVEN_HL}
\hat{H}_L^\perp(z,\psq{})\ 
=\ & \hat{G}_T(z, \psq{}),
\\ \label{EQ_SPLITTINGS_TEVEN_HTP}
\hat{H}_T^\perp(z,\psq{})\ 
=\ & C \Big[2 z^2 M^2 \Big],
}
where
\al
{
C(z,\psq{})&
\\ \non
& \equiv \frac{1-z}{12}\frac{g_\pi^2}{(2\pi)^3}  \frac{1}{(\psq{} + M^2 (1-z)^2 + z m_\pi^2)^2}.
}
%


\bibliographystyle{apsrev4-1}
\bibliography{fragment}

\begin{thebibliography}{45}%
\makeatletter
\providecommand \@ifxundefined [1]{%
 \@ifx{#1\undefined}
}%
\providecommand \@ifnum [1]{%
 \ifnum #1\expandafter \@firstoftwo
 \else \expandafter \@secondoftwo
 \fi
}%
\providecommand \@ifx [1]{%
 \ifx #1\expandafter \@firstoftwo
 \else \expandafter \@secondoftwo
 \fi
}%
\providecommand \natexlab [1]{#1}%
\providecommand \enquote  [1]{``#1''}%
\providecommand \bibnamefont  [1]{#1}%
\providecommand \bibfnamefont [1]{#1}%
\providecommand \citenamefont [1]{#1}%
\providecommand \href@noop [0]{\@secondoftwo}%
\providecommand \href [0]{\begingroup \@sanitize@url \@href}%
\providecommand \@href[1]{\@@startlink{#1}\@@href}%
\providecommand \@@href[1]{\endgroup#1\@@endlink}%
\providecommand \@sanitize@url [0]{\catcode `\\12\catcode `\$12\catcode
  `\&12\catcode `\#12\catcode `\^12\catcode `\_12\catcode `\%12\relax}%
\providecommand \@@startlink[1]{}%
\providecommand \@@endlink[0]{}%
\providecommand \url  [0]{\begingroup\@sanitize@url \@url }%
\providecommand \@url [1]{\endgroup\@href {#1}{\urlprefix }}%
\providecommand \urlprefix  [0]{URL }%
\providecommand \Eprint [0]{\href }%
\providecommand \doibase [0]{http://dx.doi.org/}%
\providecommand \selectlanguage [0]{\@gobble}%
\providecommand \bibinfo  [0]{\@secondoftwo}%
\providecommand \bibfield  [0]{\@secondoftwo}%
\providecommand \translation [1]{[#1]}%
\providecommand \BibitemOpen [0]{}%
\providecommand \bibitemStop [0]{}%
\providecommand \bibitemNoStop [0]{.\EOS\space}%
\providecommand \EOS [0]{\spacefactor3000\relax}%
\providecommand \BibitemShut  [1]{\csname bibitem#1\endcsname}%
\let\auto@bib@innerbib\@empty
\bibitem [{\citenamefont {Metz}\ and\ \citenamefont
  {Vossen}(2016)}]{Metz:2016swz}%
  \BibitemOpen
  \bibfield  {author} {\bibinfo {author} {\bibfnamefont {A.}~\bibnamefont
  {Metz}}\ and\ \bibinfo {author} {\bibfnamefont {A.}~\bibnamefont {Vossen}},\
  }\href {\doibase 10.1016/j.ppnp.2016.08.003} {\bibfield  {journal} {\bibinfo
  {journal} {Prog. Part. Nucl. Phys.}\ }\textbf {\bibinfo {volume} {91}},\
  \bibinfo {pages} {136} (\bibinfo {year} {2016})},\ \Eprint
  {http://arxiv.org/abs/1607.02521} {arXiv:1607.02521 [hep-ex]} \BibitemShut
  {NoStop}%
\bibitem [{\citenamefont {Field}\ and\ \citenamefont
  {Feynman}(1977)}]{Field:1976ve}%
  \BibitemOpen
  \bibfield  {author} {\bibinfo {author} {\bibfnamefont {R.~D.}\ \bibnamefont
  {Field}}\ and\ \bibinfo {author} {\bibfnamefont {R.~P.}\ \bibnamefont
  {Feynman}},\ }\href {\doibase 10.1103/PhysRevD.15.2590} {\bibfield  {journal}
  {\bibinfo  {journal} {Phys. Rev.}\ }\textbf {\bibinfo {volume} {D15}},\
  \bibinfo {pages} {2590} (\bibinfo {year} {1977})}\BibitemShut {NoStop}%
\bibitem [{\citenamefont {Field}\ and\ \citenamefont
  {Feynman}(1978)}]{Field:1977fa}%
  \BibitemOpen
  \bibfield  {author} {\bibinfo {author} {\bibfnamefont {R.~D.}\ \bibnamefont
  {Field}}\ and\ \bibinfo {author} {\bibfnamefont {R.~P.}\ \bibnamefont
  {Feynman}},\ }\href {\doibase 10.1016/0550-3213(78)90015-9} {\bibfield
  {journal} {\bibinfo  {journal} {Nucl. Phys.}\ }\textbf {\bibinfo {volume}
  {B136}},\ \bibinfo {pages} {1} (\bibinfo {year} {1978})}\BibitemShut
  {NoStop}%
\bibitem [{\citenamefont {Collins}(1993)}]{Collins:1992kk}%
  \BibitemOpen
  \bibfield  {author} {\bibinfo {author} {\bibfnamefont {J.~C.}\ \bibnamefont
  {Collins}},\ }\href {\doibase 10.1016/0550-3213(93)90262-N} {\bibfield
  {journal} {\bibinfo  {journal} {Nucl. Phys.}\ }\textbf {\bibinfo {volume}
  {B396}},\ \bibinfo {pages} {161} (\bibinfo {year} {1993})},\ \Eprint
  {http://arxiv.org/abs/hep-ph/9208213} {arXiv:hep-ph/9208213 [hep-ph]}
  \BibitemShut {NoStop}%
\bibitem [{\citenamefont {Kang}\ \emph {et~al.}(2016)\citenamefont {Kang},
  \citenamefont {Prokudin}, \citenamefont {Sun},\ and\ \citenamefont
  {Yuan}}]{Kang:2015msa}%
  \BibitemOpen
  \bibfield  {author} {\bibinfo {author} {\bibfnamefont {Z.-B.}\ \bibnamefont
  {Kang}}, \bibinfo {author} {\bibfnamefont {A.}~\bibnamefont {Prokudin}},
  \bibinfo {author} {\bibfnamefont {P.}~\bibnamefont {Sun}}, \ and\ \bibinfo
  {author} {\bibfnamefont {F.}~\bibnamefont {Yuan}},\ }\href {\doibase
  10.1103/PhysRevD.93.014009} {\bibfield  {journal} {\bibinfo  {journal} {Phys.
  Rev.}\ }\textbf {\bibinfo {volume} {D93}},\ \bibinfo {pages} {014009}
  (\bibinfo {year} {2016})},\ \Eprint {http://arxiv.org/abs/1505.05589}
  {arXiv:1505.05589 [hep-ph]} \BibitemShut {NoStop}%
\bibitem [{\citenamefont {Anselmino}\ \emph {et~al.}(2015)\citenamefont
  {Anselmino}, \citenamefont {Boglione}, \citenamefont {D'Alesio},
  \citenamefont {Gonzalez~Hernandez}, \citenamefont {Melis}, \citenamefont
  {Murgia},\ and\ \citenamefont {Prokudin}}]{Anselmino:2015sxa}%
  \BibitemOpen
  \bibfield  {author} {\bibinfo {author} {\bibfnamefont {M.}~\bibnamefont
  {Anselmino}}, \bibinfo {author} {\bibfnamefont {M.}~\bibnamefont {Boglione}},
  \bibinfo {author} {\bibfnamefont {U.}~\bibnamefont {D'Alesio}}, \bibinfo
  {author} {\bibfnamefont {J.~O.}\ \bibnamefont {Gonzalez~Hernandez}}, \bibinfo
  {author} {\bibfnamefont {S.}~\bibnamefont {Melis}}, \bibinfo {author}
  {\bibfnamefont {F.}~\bibnamefont {Murgia}}, \ and\ \bibinfo {author}
  {\bibfnamefont {A.}~\bibnamefont {Prokudin}},\ }\href {\doibase
  10.1103/PhysRevD.92.114023} {\bibfield  {journal} {\bibinfo  {journal} {Phys.
  Rev.}\ }\textbf {\bibinfo {volume} {D92}},\ \bibinfo {pages} {114023}
  (\bibinfo {year} {2015})},\ \Eprint {http://arxiv.org/abs/1510.05389}
  {arXiv:1510.05389 [hep-ph]} \BibitemShut {NoStop}%
\bibitem [{\citenamefont {Bacchetta}\ \emph
  {et~al.}(2008{\natexlab{a}})\citenamefont {Bacchetta}, \citenamefont
  {Gamberg}, \citenamefont {Goldstein},\ and\ \citenamefont
  {Mukherjee}}]{Bacchetta:2007wc}%
  \BibitemOpen
  \bibfield  {author} {\bibinfo {author} {\bibfnamefont {A.}~\bibnamefont
  {Bacchetta}}, \bibinfo {author} {\bibfnamefont {L.~P.}\ \bibnamefont
  {Gamberg}}, \bibinfo {author} {\bibfnamefont {G.~R.}\ \bibnamefont
  {Goldstein}}, \ and\ \bibinfo {author} {\bibfnamefont {A.}~\bibnamefont
  {Mukherjee}},\ }\href {\doibase 10.1016/j.physletb.2007.09.076} {\bibfield
  {journal} {\bibinfo  {journal} {Phys. Lett.}\ }\textbf {\bibinfo {volume}
  {B659}},\ \bibinfo {pages} {234} (\bibinfo {year} {2008}{\natexlab{a}})},\
  \Eprint {http://arxiv.org/abs/0707.3372} {arXiv:0707.3372 [hep-ph]}
  \BibitemShut {NoStop}%
\bibitem [{\citenamefont {Ito}\ \emph {et~al.}(2009)\citenamefont {Ito},
  \citenamefont {Bentz}, \citenamefont {Cloet}, \citenamefont {Thomas},\ and\
  \citenamefont {Yazaki}}]{Ito:2009zc}%
  \BibitemOpen
  \bibfield  {author} {\bibinfo {author} {\bibfnamefont {T.}~\bibnamefont
  {Ito}}, \bibinfo {author} {\bibfnamefont {W.}~\bibnamefont {Bentz}}, \bibinfo
  {author} {\bibfnamefont {I.~C.}\ \bibnamefont {Cloet}}, \bibinfo {author}
  {\bibfnamefont {A.~W.}\ \bibnamefont {Thomas}}, \ and\ \bibinfo {author}
  {\bibfnamefont {K.}~\bibnamefont {Yazaki}},\ }\href {\doibase
  10.1103/PhysRevD.80.074008} {\bibfield  {journal} {\bibinfo  {journal} {Phys.
  Rev.}\ }\textbf {\bibinfo {volume} {D80}},\ \bibinfo {pages} {074008}
  (\bibinfo {year} {2009})},\ \Eprint {http://arxiv.org/abs/0906.5362}
  {arXiv:0906.5362 [nucl-th]} \BibitemShut {NoStop}%
\bibitem [{\citenamefont {Nambu}\ and\ \citenamefont
  {Jona-Lasinio}(1961{\natexlab{a}})}]{Nambu:1961tp}%
  \BibitemOpen
  \bibfield  {author} {\bibinfo {author} {\bibfnamefont {Y.}~\bibnamefont
  {Nambu}}\ and\ \bibinfo {author} {\bibfnamefont {G.}~\bibnamefont
  {Jona-Lasinio}},\ }\href {\doibase 10.1103/PhysRev.122.345} {\bibfield
  {journal} {\bibinfo  {journal} {Phys. Rev.}\ }\textbf {\bibinfo {volume}
  {122}},\ \bibinfo {pages} {345} (\bibinfo {year}
  {1961}{\natexlab{a}})}\BibitemShut {NoStop}%
\bibitem [{\citenamefont {Nambu}\ and\ \citenamefont
  {Jona-Lasinio}(1961{\natexlab{b}})}]{Nambu:1961fr}%
  \BibitemOpen
  \bibfield  {author} {\bibinfo {author} {\bibfnamefont {Y.}~\bibnamefont
  {Nambu}}\ and\ \bibinfo {author} {\bibfnamefont {G.}~\bibnamefont
  {Jona-Lasinio}},\ }\href {\doibase 10.1103/PhysRev.124.246} {\bibfield
  {journal} {\bibinfo  {journal} {Phys. Rev.}\ }\textbf {\bibinfo {volume}
  {124}},\ \bibinfo {pages} {246} (\bibinfo {year}
  {1961}{\natexlab{b}})}\BibitemShut {NoStop}%
\bibitem [{\citenamefont {Bentz}\ \emph {et~al.}(2016)\citenamefont {Bentz},
  \citenamefont {Kotzinian}, \citenamefont {Matevosyan}, \citenamefont
  {Ninomiya}, \citenamefont {Thomas},\ and\ \citenamefont
  {Yazaki}}]{Bentz:2016rav}%
  \BibitemOpen
  \bibfield  {author} {\bibinfo {author} {\bibfnamefont {W.}~\bibnamefont
  {Bentz}}, \bibinfo {author} {\bibfnamefont {A.}~\bibnamefont {Kotzinian}},
  \bibinfo {author} {\bibfnamefont {H.~H.}\ \bibnamefont {Matevosyan}},
  \bibinfo {author} {\bibfnamefont {Y.}~\bibnamefont {Ninomiya}}, \bibinfo
  {author} {\bibfnamefont {A.~W.}\ \bibnamefont {Thomas}}, \ and\ \bibinfo
  {author} {\bibfnamefont {K.}~\bibnamefont {Yazaki}},\ }\href {\doibase
  10.1103/PhysRevD.94.034004} {\bibfield  {journal} {\bibinfo  {journal} {Phys.
  Rev.}\ }\textbf {\bibinfo {volume} {D94}},\ \bibinfo {pages} {034004}
  (\bibinfo {year} {2016})},\ \Eprint {http://arxiv.org/abs/1603.08333}
  {arXiv:1603.08333 [nucl-th]} \BibitemShut {NoStop}%
\bibitem [{\citenamefont {Andersson}\ \emph {et~al.}(1983)\citenamefont
  {Andersson}, \citenamefont {Gustafson}, \citenamefont {Ingelman},\ and\
  \citenamefont {Sjostrand}}]{Andersson:1983ia}%
  \BibitemOpen
  \bibfield  {author} {\bibinfo {author} {\bibfnamefont {B.}~\bibnamefont
  {Andersson}}, \bibinfo {author} {\bibfnamefont {G.}~\bibnamefont
  {Gustafson}}, \bibinfo {author} {\bibfnamefont {G.}~\bibnamefont {Ingelman}},
  \ and\ \bibinfo {author} {\bibfnamefont {T.}~\bibnamefont {Sjostrand}},\
  }\href {\doibase 10.1016/0370-1573(83)90080-7} {\bibfield  {journal}
  {\bibinfo  {journal} {Phys. Rept.}\ }\textbf {\bibinfo {volume} {97}},\
  \bibinfo {pages} {31} (\bibinfo {year} {1983})}\BibitemShut {NoStop}%
\bibitem [{\citenamefont {Sjostrand}\ \emph {et~al.}(2008)\citenamefont
  {Sjostrand}, \citenamefont {Mrenna},\ and\ \citenamefont
  {Skands}}]{Sjostrand:2007gs}%
  \BibitemOpen
  \bibfield  {author} {\bibinfo {author} {\bibfnamefont {T.}~\bibnamefont
  {Sjostrand}}, \bibinfo {author} {\bibfnamefont {S.}~\bibnamefont {Mrenna}}, \
  and\ \bibinfo {author} {\bibfnamefont {P.~Z.}\ \bibnamefont {Skands}},\
  }\href {\doibase 10.1016/j.cpc.2008.01.036} {\bibfield  {journal} {\bibinfo
  {journal} {Comput. Phys. Commun.}\ }\textbf {\bibinfo {volume} {178}},\
  \bibinfo {pages} {852} (\bibinfo {year} {2008})},\ \Eprint
  {http://arxiv.org/abs/0710.3820} {arXiv:0710.3820 [hep-ph]} \BibitemShut
  {NoStop}%
\bibitem [{\citenamefont {Ingelman}\ \emph {et~al.}(1997)\citenamefont
  {Ingelman}, \citenamefont {Edin},\ and\ \citenamefont
  {Rathsman}}]{Ingelman:1996mq}%
  \BibitemOpen
  \bibfield  {author} {\bibinfo {author} {\bibfnamefont {G.}~\bibnamefont
  {Ingelman}}, \bibinfo {author} {\bibfnamefont {A.}~\bibnamefont {Edin}}, \
  and\ \bibinfo {author} {\bibfnamefont {J.}~\bibnamefont {Rathsman}},\ }\href
  {\doibase 10.1016/S0010-4655(96)00157-9} {\bibfield  {journal} {\bibinfo
  {journal} {Comput.Phys.Commun.}\ }\textbf {\bibinfo {volume} {101}},\
  \bibinfo {pages} {108} (\bibinfo {year} {1997})},\ \Eprint
  {http://arxiv.org/abs/hep-ph/9605286} {arXiv:hep-ph/9605286 [hep-ph]}
  \BibitemShut {NoStop}%
\bibitem [{\citenamefont {Sivers}(1990)}]{Sivers:1989cc}%
  \BibitemOpen
  \bibfield  {author} {\bibinfo {author} {\bibfnamefont {D.~W.}\ \bibnamefont
  {Sivers}},\ }\href {\doibase 10.1103/PhysRevD.41.83} {\bibfield  {journal}
  {\bibinfo  {journal} {Phys.Rev.}\ }\textbf {\bibinfo {volume} {D41}},\
  \bibinfo {pages} {83} (\bibinfo {year} {1990})}\BibitemShut {NoStop}%
\bibitem [{\citenamefont {Kotzinian}\ \emph
  {et~al.}(2014{\natexlab{a}})\citenamefont {Kotzinian}, \citenamefont
  {Matevosyan},\ and\ \citenamefont {Thomas}}]{Kotzinian:2014lsa}%
  \BibitemOpen
  \bibfield  {author} {\bibinfo {author} {\bibfnamefont {A.}~\bibnamefont
  {Kotzinian}}, \bibinfo {author} {\bibfnamefont {H.~H.}\ \bibnamefont
  {Matevosyan}}, \ and\ \bibinfo {author} {\bibfnamefont {A.~W.}\ \bibnamefont
  {Thomas}},\ }\href {\doibase 10.1103/PhysRevLett.113.062003} {\bibfield
  {journal} {\bibinfo  {journal} {Phys.Rev.Lett.}\ }\textbf {\bibinfo {volume}
  {113}},\ \bibinfo {pages} {062003} (\bibinfo {year} {2014}{\natexlab{a}})},\
  \Eprint {http://arxiv.org/abs/1403.5562} {arXiv:1403.5562 [hep-ph]}
  \BibitemShut {NoStop}%
\bibitem [{\citenamefont {Kotzinian}\ \emph
  {et~al.}(2014{\natexlab{b}})\citenamefont {Kotzinian}, \citenamefont
  {Matevosyan},\ and\ \citenamefont {Thomas}}]{Kotzinian:2014gza}%
  \BibitemOpen
  \bibfield  {author} {\bibinfo {author} {\bibfnamefont {A.}~\bibnamefont
  {Kotzinian}}, \bibinfo {author} {\bibfnamefont {H.~H.}\ \bibnamefont
  {Matevosyan}}, \ and\ \bibinfo {author} {\bibfnamefont {A.~W.}\ \bibnamefont
  {Thomas}},\ }\href {\doibase 10.1103/PhysRevD.90.074006} {\bibfield
  {journal} {\bibinfo  {journal} {Phys.Rev.}\ }\textbf {\bibinfo {volume}
  {D90}},\ \bibinfo {pages} {074006} (\bibinfo {year} {2014}{\natexlab{b}})},\
  \Eprint {http://arxiv.org/abs/1405.5059} {arXiv:1405.5059 [hep-ph]}
  \BibitemShut {NoStop}%
\bibitem [{\citenamefont {Matevosyan}\ \emph {et~al.}(2015)\citenamefont
  {Matevosyan}, \citenamefont {Kotzinian}, \citenamefont {Aschenauer},
  \citenamefont {Avakian},\ and\ \citenamefont {Thomas}}]{Matevosyan:2015gwa}%
  \BibitemOpen
  \bibfield  {author} {\bibinfo {author} {\bibfnamefont {H.~H.}\ \bibnamefont
  {Matevosyan}}, \bibinfo {author} {\bibfnamefont {A.}~\bibnamefont
  {Kotzinian}}, \bibinfo {author} {\bibfnamefont {E.-C.}\ \bibnamefont
  {Aschenauer}}, \bibinfo {author} {\bibfnamefont {H.}~\bibnamefont {Avakian}},
  \ and\ \bibinfo {author} {\bibfnamefont {A.~W.}\ \bibnamefont {Thomas}},\
  }\href {\doibase 10.1103/PhysRevD.92.054028} {\bibfield  {journal} {\bibinfo
  {journal} {Phys. Rev.}\ }\textbf {\bibinfo {volume} {D92}},\ \bibinfo {pages}
  {054028} (\bibinfo {year} {2015})},\ \Eprint
  {http://arxiv.org/abs/1502.02669} {arXiv:1502.02669 [hep-ph]} \BibitemShut
  {NoStop}%
\bibitem [{\citenamefont {Artru}\ and\ \citenamefont
  {Belghobsi}(2014)}]{Artru:2014abc}%
  \BibitemOpen
  \bibfield  {author} {\bibinfo {author} {\bibfnamefont {X.}~\bibnamefont
  {Artru}}\ and\ \bibinfo {author} {\bibfnamefont {Z.}~\bibnamefont
  {Belghobsi}},\ }in\ \href@noop {} {\emph {\bibinfo {booktitle} {Proc. of XV
  Advanced Research Workshop on High Energy Spin Physics (DSPIN 2013)}}},\
  \bibinfo {editor} {edited by\ \bibinfo {editor} {\bibfnamefont {A.~V.}\
  \bibnamefont {Efremov}}\ and\ \bibinfo {editor} {\bibfnamefont {S.~V.}\
  \bibnamefont {Goloskokov}}}\ (\bibinfo  {publisher} {Russian Federation,
  Dubna},\ \bibinfo {year} {2014})\ pp.\ \bibinfo {pages} {33--40}\BibitemShut
  {NoStop}%
\bibitem [{\citenamefont {Kotzinian}(2005)}]{Kotzinian:2004xq}%
  \BibitemOpen
  \bibfield  {author} {\bibinfo {author} {\bibfnamefont {A.}~\bibnamefont
  {Kotzinian}},\ }\href {\doibase 10.1140/epjc/s2005-02319-5} {\bibfield
  {journal} {\bibinfo  {journal} {Eur.Phys.J.}\ }\textbf {\bibinfo {volume}
  {C44}},\ \bibinfo {pages} {211} (\bibinfo {year} {2005})},\ \Eprint
  {http://arxiv.org/abs/hep-ph/0410093} {arXiv:hep-ph/0410093 [hep-ph]}
  \BibitemShut {NoStop}%
\bibitem [{\citenamefont {Matevosyan}\ \emph
  {et~al.}(2011{\natexlab{a}})\citenamefont {Matevosyan}, \citenamefont
  {Thomas},\ and\ \citenamefont {Bentz}}]{Matevosyan:2010hh}%
  \BibitemOpen
  \bibfield  {author} {\bibinfo {author} {\bibfnamefont {H.~H.}\ \bibnamefont
  {Matevosyan}}, \bibinfo {author} {\bibfnamefont {A.~W.}\ \bibnamefont
  {Thomas}}, \ and\ \bibinfo {author} {\bibfnamefont {W.}~\bibnamefont
  {Bentz}},\ }\href {\doibase 10.1103/PhysRevD.83.074003} {\bibfield  {journal}
  {\bibinfo  {journal} {Phys.Rev.}\ }\textbf {\bibinfo {volume} {D83}},\
  \bibinfo {pages} {074003} (\bibinfo {year} {2011}{\natexlab{a}})},\ \Eprint
  {http://arxiv.org/abs/1011.1052} {arXiv:1011.1052 [hep-ph]} \BibitemShut
  {NoStop}%
\bibitem [{\citenamefont {Matevosyan}\ \emph
  {et~al.}(2011{\natexlab{b}})\citenamefont {Matevosyan}, \citenamefont
  {Thomas},\ and\ \citenamefont {Bentz}}]{Matevosyan:2011ey}%
  \BibitemOpen
  \bibfield  {author} {\bibinfo {author} {\bibfnamefont {H.~H.}\ \bibnamefont
  {Matevosyan}}, \bibinfo {author} {\bibfnamefont {A.~W.}\ \bibnamefont
  {Thomas}}, \ and\ \bibinfo {author} {\bibfnamefont {W.}~\bibnamefont
  {Bentz}},\ }\href {\doibase 10.1103/PhysRevD.83.114010,
  10.1103/PhysRevD.86.059904} {\bibfield  {journal} {\bibinfo  {journal}
  {Phys.Rev.}\ }\textbf {\bibinfo {volume} {D83}},\ \bibinfo {pages} {114010}
  (\bibinfo {year} {2011}{\natexlab{b}})},\ \Eprint
  {http://arxiv.org/abs/1103.3085} {arXiv:1103.3085 [hep-ph]} \BibitemShut
  {NoStop}%
\bibitem [{\citenamefont {Matevosyan}\ \emph
  {et~al.}(2012{\natexlab{a}})\citenamefont {Matevosyan}, \citenamefont
  {Thomas},\ and\ \citenamefont {Bentz}}]{PhysRevD.86.059904}%
  \BibitemOpen
  \bibfield  {author} {\bibinfo {author} {\bibfnamefont {H.~H.}\ \bibnamefont
  {Matevosyan}}, \bibinfo {author} {\bibfnamefont {A.~W.}\ \bibnamefont
  {Thomas}}, \ and\ \bibinfo {author} {\bibfnamefont {W.}~\bibnamefont
  {Bentz}},\ }\href {\doibase 10.1103/PhysRevD.86.059904} {\bibfield  {journal}
  {\bibinfo  {journal} {Phys. Rev. D}\ }\textbf {\bibinfo {volume} {86}},\
  \bibinfo {pages} {059904(E)} (\bibinfo {year}
  {2012}{\natexlab{a}})}\BibitemShut {NoStop}%
\bibitem [{\citenamefont {Matevosyan}\ \emph
  {et~al.}(2012{\natexlab{b}})\citenamefont {Matevosyan}, \citenamefont
  {Bentz}, \citenamefont {Cloet},\ and\ \citenamefont
  {Thomas}}]{Matevosyan:2011vj}%
  \BibitemOpen
  \bibfield  {author} {\bibinfo {author} {\bibfnamefont {H.~H.}\ \bibnamefont
  {Matevosyan}}, \bibinfo {author} {\bibfnamefont {W.}~\bibnamefont {Bentz}},
  \bibinfo {author} {\bibfnamefont {I.~C.}\ \bibnamefont {Cloet}}, \ and\
  \bibinfo {author} {\bibfnamefont {A.~W.}\ \bibnamefont {Thomas}},\ }\href
  {\doibase 10.1103/PhysRevD.85.014021} {\bibfield  {journal} {\bibinfo
  {journal} {Phys.Rev.}\ }\textbf {\bibinfo {volume} {D85}},\ \bibinfo {pages}
  {014021} (\bibinfo {year} {2012}{\natexlab{b}})},\ \Eprint
  {http://arxiv.org/abs/1111.1740} {arXiv:1111.1740 [hep-ph]} \BibitemShut
  {NoStop}%
\bibitem [{\citenamefont {Matevosyan}\ \emph
  {et~al.}(2014{\natexlab{a}})\citenamefont {Matevosyan}, \citenamefont
  {Thomas},\ and\ \citenamefont {Bentz}}]{Matevosyan:2013nla}%
  \BibitemOpen
  \bibfield  {author} {\bibinfo {author} {\bibfnamefont {H.~H.}\ \bibnamefont
  {Matevosyan}}, \bibinfo {author} {\bibfnamefont {A.~W.}\ \bibnamefont
  {Thomas}}, \ and\ \bibinfo {author} {\bibfnamefont {W.}~\bibnamefont
  {Bentz}},\ }\bibfield  {booktitle} {\emph {\bibinfo {booktitle}
  {{Proceedings, 25th International Nuclear Physics Conference (INPC 2013)}}},\
  }\href {\doibase 10.1051/epjconf/20146606014} {\bibfield  {journal} {\bibinfo
   {journal} {EPJ Web Conf.}\ }\textbf {\bibinfo {volume} {66}},\ \bibinfo
  {pages} {06014} (\bibinfo {year} {2014}{\natexlab{a}})},\ \Eprint
  {http://arxiv.org/abs/1307.8125} {arXiv:1307.8125 [hep-ph]} \BibitemShut
  {NoStop}%
\bibitem [{\citenamefont {Matevosyan}\ \emph {et~al.}(2013)\citenamefont
  {Matevosyan}, \citenamefont {Thomas},\ and\ \citenamefont
  {Bentz}}]{Matevosyan:2013aka}%
  \BibitemOpen
  \bibfield  {author} {\bibinfo {author} {\bibfnamefont {H.~H.}\ \bibnamefont
  {Matevosyan}}, \bibinfo {author} {\bibfnamefont {A.~W.}\ \bibnamefont
  {Thomas}}, \ and\ \bibinfo {author} {\bibfnamefont {W.}~\bibnamefont
  {Bentz}},\ }\href {\doibase 10.1103/PhysRevD.88.094022} {\bibfield  {journal}
  {\bibinfo  {journal} {Phys.Rev.}\ }\textbf {\bibinfo {volume} {D88}},\
  \bibinfo {pages} {094022} (\bibinfo {year} {2013})},\ \Eprint
  {http://arxiv.org/abs/1310.1917} {arXiv:1310.1917 [hep-ph]} \BibitemShut
  {NoStop}%
\bibitem [{\citenamefont {Casey}\ \emph
  {et~al.}(2012{\natexlab{a}})\citenamefont {Casey}, \citenamefont
  {Matevosyan},\ and\ \citenamefont {Thomas}}]{Casey:2012ux}%
  \BibitemOpen
  \bibfield  {author} {\bibinfo {author} {\bibfnamefont {A.}~\bibnamefont
  {Casey}}, \bibinfo {author} {\bibfnamefont {H.~H.}\ \bibnamefont
  {Matevosyan}}, \ and\ \bibinfo {author} {\bibfnamefont {A.~W.}\ \bibnamefont
  {Thomas}},\ }\href {\doibase 10.1103/PhysRevD.85.114049} {\bibfield
  {journal} {\bibinfo  {journal} {Phys. Rev. D}\ }\textbf {\bibinfo {volume}
  {85}},\ \bibinfo {pages} {114049} (\bibinfo {year} {2012}{\natexlab{a}})},\
  \Eprint {http://arxiv.org/abs/1202.4036} {arXiv:1202.4036 [hep-ph]}
  \BibitemShut {NoStop}%
\bibitem [{\citenamefont {Casey}\ \emph
  {et~al.}(2012{\natexlab{b}})\citenamefont {Casey}, \citenamefont {Cloet},
  \citenamefont {Matevosyan},\ and\ \citenamefont {Thomas}}]{Casey:2012hg}%
  \BibitemOpen
  \bibfield  {author} {\bibinfo {author} {\bibfnamefont {A.}~\bibnamefont
  {Casey}}, \bibinfo {author} {\bibfnamefont {I.~C.}\ \bibnamefont {Cloet}},
  \bibinfo {author} {\bibfnamefont {H.~H.}\ \bibnamefont {Matevosyan}}, \ and\
  \bibinfo {author} {\bibfnamefont {A.~W.}\ \bibnamefont {Thomas}},\ }\href
  {\doibase 10.1103/PhysRevD.86.114018} {\bibfield  {journal} {\bibinfo
  {journal} {Phys.Rev.}\ }\textbf {\bibinfo {volume} {D86}},\ \bibinfo {pages}
  {114018} (\bibinfo {year} {2012}{\natexlab{b}})},\ \Eprint
  {http://arxiv.org/abs/1207.4267} {arXiv:1207.4267 [hep-ph]} \BibitemShut
  {NoStop}%
\bibitem [{\citenamefont {Matevosyan}\ \emph
  {et~al.}(2012{\natexlab{c}})\citenamefont {Matevosyan}, \citenamefont
  {Thomas},\ and\ \citenamefont {Bentz}}]{Matevosyan:2012ga}%
  \BibitemOpen
  \bibfield  {author} {\bibinfo {author} {\bibfnamefont {H.~H.}\ \bibnamefont
  {Matevosyan}}, \bibinfo {author} {\bibfnamefont {A.~W.}\ \bibnamefont
  {Thomas}}, \ and\ \bibinfo {author} {\bibfnamefont {W.}~\bibnamefont
  {Bentz}},\ }\href {\doibase 10.1103/PhysRevD.86.034025} {\bibfield  {journal}
  {\bibinfo  {journal} {Phys.Rev.}\ }\textbf {\bibinfo {volume} {D86}},\
  \bibinfo {pages} {034025} (\bibinfo {year} {2012}{\natexlab{c}})},\ \Eprint
  {http://arxiv.org/abs/1205.5813} {arXiv:1205.5813 [hep-ph]} \BibitemShut
  {NoStop}%
\bibitem [{\citenamefont {Matevosyan}\ \emph
  {et~al.}(2012{\natexlab{d}})\citenamefont {Matevosyan}, \citenamefont
  {Thomas},\ and\ \citenamefont {Bentz}}]{Matevosyan:2012ms}%
  \BibitemOpen
  \bibfield  {author} {\bibinfo {author} {\bibfnamefont {H.~H.}\ \bibnamefont
  {Matevosyan}}, \bibinfo {author} {\bibfnamefont {A.~W.}\ \bibnamefont
  {Thomas}}, \ and\ \bibinfo {author} {\bibfnamefont {W.}~\bibnamefont
  {Bentz}},\ }\href {\doibase 10.1088/1742-6596/403/1/012042} {\bibfield
  {journal} {\bibinfo  {journal} {J.Phys.Conf.Ser.}\ }\textbf {\bibinfo
  {volume} {403}},\ \bibinfo {pages} {012042} (\bibinfo {year}
  {2012}{\natexlab{d}})},\ \Eprint {http://arxiv.org/abs/1207.0812}
  {arXiv:1207.0812 [hep-ph]} \BibitemShut {NoStop}%
\bibitem [{\citenamefont {Matevosyan}\ \emph
  {et~al.}(2014{\natexlab{b}})\citenamefont {Matevosyan}, \citenamefont
  {Kotzinian},\ and\ \citenamefont {Thomas}}]{Matevosyan:2013eia}%
  \BibitemOpen
  \bibfield  {author} {\bibinfo {author} {\bibfnamefont {H.~H.}\ \bibnamefont
  {Matevosyan}}, \bibinfo {author} {\bibfnamefont {A.}~\bibnamefont
  {Kotzinian}}, \ and\ \bibinfo {author} {\bibfnamefont {A.~W.}\ \bibnamefont
  {Thomas}},\ }\href {\doibase 10.1016/j.physletb.2014.02.040} {\bibfield
  {journal} {\bibinfo  {journal} {Phys.Lett.}\ }\textbf {\bibinfo {volume}
  {B731}},\ \bibinfo {pages} {208} (\bibinfo {year} {2014}{\natexlab{b}})},\
  \Eprint {http://arxiv.org/abs/1312.4556} {arXiv:1312.4556 [hep-ph]}
  \BibitemShut {NoStop}%
\bibitem [{\citenamefont {Matevosyan}\ \emph
  {et~al.}(2012{\natexlab{e}})\citenamefont {Matevosyan}, \citenamefont
  {Thomas},\ and\ \citenamefont {Bentz}}]{Matevosyan:2012ed}%
  \BibitemOpen
  \bibfield  {author} {\bibinfo {author} {\bibfnamefont {H.~H.}\ \bibnamefont
  {Matevosyan}}, \bibinfo {author} {\bibfnamefont {A.~W.}\ \bibnamefont
  {Thomas}}, \ and\ \bibinfo {author} {\bibfnamefont {W.}~\bibnamefont
  {Bentz}},\ }\href@noop {} {\enquote {\bibinfo {title} {{Higher Order Collins
  Modulations in Transversely Polarized Quark Fragmentation}},}\ } (\bibinfo
  {year} {2012}{\natexlab{e}}),\ \Eprint {http://arxiv.org/abs/1207.1433}
  {arXiv:1207.1433 [hep-ph]} \BibitemShut {NoStop}%
\bibitem [{\citenamefont {Adolph}\ \emph {et~al.}(2014)\citenamefont {Adolph}
  \emph {et~al.}}]{Adolph:2014fjw}%
  \BibitemOpen
  \bibfield  {author} {\bibinfo {author} {\bibfnamefont {C.}~\bibnamefont
  {Adolph}} \emph {et~al.} (\bibinfo {collaboration} {COMPASS Collaboration}),\
  }\href {\doibase 10.1016/j.physletb.2014.06.080} {\bibfield  {journal}
  {\bibinfo  {journal} {Phys.Lett.}\ }\textbf {\bibinfo {volume} {B736}},\
  \bibinfo {pages} {124} (\bibinfo {year} {2014})},\ \Eprint
  {http://arxiv.org/abs/1401.7873} {arXiv:1401.7873 [hep-ex]} \BibitemShut
  {NoStop}%
\bibitem [{\citenamefont {Berestetskii}\ \emph {et~al.}(1982)\citenamefont
  {Berestetskii}, \citenamefont {Lifshitz},\ and\ \citenamefont
  {Pitaevskii}}]{Berestetsky:1982aq}%
  \BibitemOpen
  \bibfield  {author} {\bibinfo {author} {\bibfnamefont {V.~B.}\ \bibnamefont
  {Berestetskii}}, \bibinfo {author} {\bibfnamefont {E.~M.}\ \bibnamefont
  {Lifshitz}}, \ and\ \bibinfo {author} {\bibfnamefont {L.~P.}\ \bibnamefont
  {Pitaevskii}},\ }\href
  {http://www-spires.fnal.gov/spires/find/books/www?cl=QC680.B42} {\emph
  {\bibinfo {title} {{QUANTUM ELECTRODYNAMICS}}}},\ \bibinfo {series} {Course
  of Theoretical Physics}, Vol.~\bibinfo {volume} {4}\ (\bibinfo  {publisher}
  {Pergamon Press},\ \bibinfo {address} {Oxford},\ \bibinfo {year}
  {1982})\BibitemShut {NoStop}%
\bibitem [{\citenamefont {Kotzinian}(1995)}]{Kotzinian:1994dv}%
  \BibitemOpen
  \bibfield  {author} {\bibinfo {author} {\bibfnamefont {A.}~\bibnamefont
  {Kotzinian}},\ }\href {\doibase 10.1016/0550-3213(95)00098-D} {\bibfield
  {journal} {\bibinfo  {journal} {Nucl.Phys.}\ }\textbf {\bibinfo {volume}
  {B441}},\ \bibinfo {pages} {234} (\bibinfo {year} {1995})},\ \Eprint
  {http://arxiv.org/abs/hep-ph/9412283} {arXiv:hep-ph/9412283 [hep-ph]}
  \BibitemShut {NoStop}%
\bibitem [{\citenamefont {Meissner}\ \emph {et~al.}(2010)\citenamefont
  {Meissner}, \citenamefont {Metz},\ and\ \citenamefont
  {Pitonyak}}]{Meissner:2010cc}%
  \BibitemOpen
  \bibfield  {author} {\bibinfo {author} {\bibfnamefont {S.}~\bibnamefont
  {Meissner}}, \bibinfo {author} {\bibfnamefont {A.}~\bibnamefont {Metz}}, \
  and\ \bibinfo {author} {\bibfnamefont {D.}~\bibnamefont {Pitonyak}},\ }\href
  {\doibase 10.1016/j.physletb.2010.05.037} {\bibfield  {journal} {\bibinfo
  {journal} {Phys.Lett.}\ }\textbf {\bibinfo {volume} {B690}},\ \bibinfo
  {pages} {296} (\bibinfo {year} {2010})},\ \Eprint
  {http://arxiv.org/abs/1002.4393} {arXiv:1002.4393 [hep-ph]} \BibitemShut
  {NoStop}%
\bibitem [{\citenamefont {Amrath}\ \emph {et~al.}(2005)\citenamefont {Amrath},
  \citenamefont {Bacchetta},\ and\ \citenamefont {Metz}}]{Amrath:2005gv}%
  \BibitemOpen
  \bibfield  {author} {\bibinfo {author} {\bibfnamefont {D.}~\bibnamefont
  {Amrath}}, \bibinfo {author} {\bibfnamefont {A.}~\bibnamefont {Bacchetta}}, \
  and\ \bibinfo {author} {\bibfnamefont {A.}~\bibnamefont {Metz}},\ }\href
  {\doibase 10.1103/PhysRevD.71.114018} {\bibfield  {journal} {\bibinfo
  {journal} {Phys. Rev.}\ }\textbf {\bibinfo {volume} {D71}},\ \bibinfo {pages}
  {114018} (\bibinfo {year} {2005})},\ \Eprint
  {http://arxiv.org/abs/hep-ph/0504124} {arXiv:hep-ph/0504124} \BibitemShut
  {NoStop}%
\bibitem [{\citenamefont {Bacchetta}\ \emph {et~al.}(2000)\citenamefont
  {Bacchetta}, \citenamefont {Boglione}, \citenamefont {Henneman},\ and\
  \citenamefont {Mulders}}]{Bacchetta:1999kz}%
  \BibitemOpen
  \bibfield  {author} {\bibinfo {author} {\bibfnamefont {A.}~\bibnamefont
  {Bacchetta}}, \bibinfo {author} {\bibfnamefont {M.}~\bibnamefont {Boglione}},
  \bibinfo {author} {\bibfnamefont {A.}~\bibnamefont {Henneman}}, \ and\
  \bibinfo {author} {\bibfnamefont {P.~J.}\ \bibnamefont {Mulders}},\ }\href
  {\doibase 10.1103/PhysRevLett.85.712} {\bibfield  {journal} {\bibinfo
  {journal} {Phys. Rev. Lett.}\ }\textbf {\bibinfo {volume} {85}},\ \bibinfo
  {pages} {712} (\bibinfo {year} {2000})},\ \Eprint
  {http://arxiv.org/abs/hep-ph/9912490} {arXiv:hep-ph/9912490 [hep-ph]}
  \BibitemShut {NoStop}%
\bibitem [{\citenamefont {Kotzinian}(2008)}]{Kotzinian:2008fe}%
  \BibitemOpen
  \bibfield  {author} {\bibinfo {author} {\bibfnamefont {A.}~\bibnamefont
  {Kotzinian}},\ }in\ \href
  {http://inspirehep.net/record/788980/files/arXiv:0806.3804.pdf} {\emph
  {\bibinfo {booktitle} {{Transversity 2008: 2nd International Workshop on
  Transverse Polarization Phenomena in Hard Processes Ferrara, Italy, May
  28-31, 2008}}}}\ (\bibinfo {year} {2008})\ \Eprint
  {http://arxiv.org/abs/0806.3804} {arXiv:0806.3804 [hep-ph]} \BibitemShut
  {NoStop}%
\bibitem [{\citenamefont {Ellis}\ \emph {et~al.}(2009)\citenamefont {Ellis},
  \citenamefont {Hwang},\ and\ \citenamefont {Kotzinian}}]{Ellis:2008in}%
  \BibitemOpen
  \bibfield  {author} {\bibinfo {author} {\bibfnamefont {J.~R.}\ \bibnamefont
  {Ellis}}, \bibinfo {author} {\bibfnamefont {D.~S.}\ \bibnamefont {Hwang}}, \
  and\ \bibinfo {author} {\bibfnamefont {A.}~\bibnamefont {Kotzinian}},\ }\href
  {\doibase 10.1103/PhysRevD.80.074033} {\bibfield  {journal} {\bibinfo
  {journal} {Phys. Rev.}\ }\textbf {\bibinfo {volume} {D80}},\ \bibinfo {pages}
  {074033} (\bibinfo {year} {2009})},\ \Eprint {http://arxiv.org/abs/0808.1567}
  {arXiv:0808.1567 [hep-ph]} \BibitemShut {NoStop}%
\bibitem [{\citenamefont {Bacchetta}\ \emph
  {et~al.}(2008{\natexlab{b}})\citenamefont {Bacchetta}, \citenamefont
  {Conti},\ and\ \citenamefont {Radici}}]{Bacchetta:2008af}%
  \BibitemOpen
  \bibfield  {author} {\bibinfo {author} {\bibfnamefont {A.}~\bibnamefont
  {Bacchetta}}, \bibinfo {author} {\bibfnamefont {F.}~\bibnamefont {Conti}}, \
  and\ \bibinfo {author} {\bibfnamefont {M.}~\bibnamefont {Radici}},\ }\href
  {\doibase 10.1103/PhysRevD.78.074010} {\bibfield  {journal} {\bibinfo
  {journal} {Phys. Rev.}\ }\textbf {\bibinfo {volume} {D78}},\ \bibinfo {pages}
  {074010} (\bibinfo {year} {2008}{\natexlab{b}})},\ \Eprint
  {http://arxiv.org/abs/0807.0323} {arXiv:0807.0323 [hep-ph]} \BibitemShut
  {NoStop}%
\bibitem [{\citenamefont {Pasquini}\ and\ \citenamefont
  {Schweitzer}(2011)}]{Pasquini:2011tk}%
  \BibitemOpen
  \bibfield  {author} {\bibinfo {author} {\bibfnamefont {B.}~\bibnamefont
  {Pasquini}}\ and\ \bibinfo {author} {\bibfnamefont {P.}~\bibnamefont
  {Schweitzer}},\ }\href {\doibase 10.1103/PhysRevD.83.114044} {\bibfield
  {journal} {\bibinfo  {journal} {Phys. Rev.}\ }\textbf {\bibinfo {volume}
  {D83}},\ \bibinfo {pages} {114044} (\bibinfo {year} {2011})},\ \Eprint
  {http://arxiv.org/abs/1103.5977} {arXiv:1103.5977 [hep-ph]} \BibitemShut
  {NoStop}%
\bibitem [{\citenamefont {Bahr}\ \emph {et~al.}(2008)\citenamefont {Bahr} \emph
  {et~al.}}]{Bahr:2008pv}%
  \BibitemOpen
  \bibfield  {author} {\bibinfo {author} {\bibfnamefont {M.}~\bibnamefont
  {Bahr}} \emph {et~al.},\ }\href {\doibase 10.1140/epjc/s10052-008-0798-9}
  {\bibfield  {journal} {\bibinfo  {journal} {Eur. Phys. J.}\ }\textbf
  {\bibinfo {volume} {C58}},\ \bibinfo {pages} {639} (\bibinfo {year}
  {2008})},\ \Eprint {http://arxiv.org/abs/0803.0883} {arXiv:0803.0883
  [hep-ph]} \BibitemShut {NoStop}%
\bibitem [{\citenamefont {Gleisberg}\ \emph {et~al.}(2009)\citenamefont
  {Gleisberg}, \citenamefont {Hoeche}, \citenamefont {Krauss}, \citenamefont
  {Schonherr}, \citenamefont {Schumann}, \citenamefont {Siegert},\ and\
  \citenamefont {Winter}}]{Gleisberg:2008ta}%
  \BibitemOpen
  \bibfield  {author} {\bibinfo {author} {\bibfnamefont {T.}~\bibnamefont
  {Gleisberg}}, \bibinfo {author} {\bibfnamefont {S.}~\bibnamefont {Hoeche}},
  \bibinfo {author} {\bibfnamefont {F.}~\bibnamefont {Krauss}}, \bibinfo
  {author} {\bibfnamefont {M.}~\bibnamefont {Schonherr}}, \bibinfo {author}
  {\bibfnamefont {S.}~\bibnamefont {Schumann}}, \bibinfo {author}
  {\bibfnamefont {F.}~\bibnamefont {Siegert}}, \ and\ \bibinfo {author}
  {\bibfnamefont {J.}~\bibnamefont {Winter}},\ }\href {\doibase
  10.1088/1126-6708/2009/02/007} {\bibfield  {journal} {\bibinfo  {journal}
  {JHEP}\ }\textbf {\bibinfo {volume} {02}},\ \bibinfo {pages} {007} (\bibinfo
  {year} {2009})},\ \Eprint {http://arxiv.org/abs/0811.4622} {arXiv:0811.4622
  [hep-ph]} \BibitemShut {NoStop}%
\bibitem [{\citenamefont {Adkins}\ and\ \citenamefont
  {Drachenberg}(2016)}]{Adkins:2016uxv}%
  \BibitemOpen
  \bibfield  {author} {\bibinfo {author} {\bibfnamefont {J.~K.}\ \bibnamefont
  {Adkins}}\ and\ \bibinfo {author} {\bibfnamefont {J.~L.}\ \bibnamefont
  {Drachenberg}} (\bibinfo {collaboration} {STAR}),\ }\bibfield  {booktitle}
  {\emph {\bibinfo {booktitle} {{Proceedings, 21st International Symposium on
  Spin Physics (SPIN 2014): Beijing, China, October 20-24, 2014}}},\ }\href
  {\doibase 10.1142/S2010194516600405} {\bibfield  {journal} {\bibinfo
  {journal} {Int. J. Mod. Phys. Conf. Ser.}\ }\textbf {\bibinfo {volume}
  {40}},\ \bibinfo {pages} {1660040} (\bibinfo {year} {2016})}\BibitemShut
  {NoStop}%
\end{thebibliography}%

\end{document}